\documentclass[usenatbib,iop,numberedappendix]{aeb_emulateapj_2010}

\usepackage{amsmath}
\usepackage{amssymb}
\usepackage{xspace}
\usepackage[normalem]{ulem}

\usepackage{natbib}

\usepackage[usenames,dvips]{color}

\def\del#1{{}}

\def\GHz{{\rm GHz}} 

\def\m{{\rm m}} 
\def\mm{{\rm m}\m} 
\def\cm{{\rm c}\m} 

\def\Jy{{\rm Jy}} 

\def\mas{{\rm mas}} 

\renewcommand{\d}{{\rm d}}

\def\Rg{{r_{\rm g}}}

\def\BIC{{\rm BIC}}
\def\AIC{{\rm AIC}}

\newcommand\bmath[1] {\mbox{\boldmath$\rm #1$}}

\newcommand\p{\bmath{p}}
\newcommand\q{\bmath{q}}

\begin{document}

\title{
Testing the No-Hair Theorem with Event Horizon Telescope Observations of Sagittarius A*
}

\author{
Avery E.~Broderick,\altaffilmark{1,2}
Tim Johannsen,\altaffilmark{1,2,3}\footnote{CITA National Fellow}
Abraham Loeb,\altaffilmark{4}
and Dimitrios Psaltis\altaffilmark{5}
}
\altaffiltext{1}{Perimeter Institute for Theoretical Physics, 31 Caroline Street North, Waterloo, ON, N2L 2Y5, Canada}
\altaffiltext{2}{Department of Physics and Astronomy, University of Waterloo, 200 University Avenue West, Waterloo, ON, N2L 3G1, Canada}
\altaffiltext{3}{Canadian Institute for Theoretical Astrophysics, 60 St.~George Street, Toronto, ON M5S 3H8, Canada}
\altaffiltext{4}{Institute for Theory and Computation, Harvard University, Center for Astrophysics, 60 Garden Street, Cambridge, MA 02138, USA}
\altaffiltext{5}{Astronomy and Physics Departments, University of Arizona, 933 North Cherry Street, Tucson, AZ 85721, USA}

\shorttitle{Testing the No Hair Theorem with Sgr A*}
\shortauthors{Broderick et al.}

\begin{abstract}
The advent of the Event Horizon Telescope (EHT), a millimeter-wave
very-long baseline interferometric array, has enabled
spatially-resolved studies of the sub-horizon-scale structure for a
handful of supermassive black holes. Among these, the supermassive
black hole at the center of the Milky Way, Sagittarius A* (Sgr~A*),
presents the largest angular cross section.  Thus far, these studies
have focused upon measurements of the black hole spin and the
validation of low-luminosity accretion models.  However, a critical
input into the analysis of EHT data is the structure of the black hole
spacetime, and thus these observations provide the novel opportunity
to test the applicability of the Kerr metric to astrophysical black
holes.  Here we present the first simulated images of a radiatively
inefficient accretion flow (RIAF) around Sgr~A* employing a quasi-Kerr
metric that contains an independent quadrupole moment in addition to
the mass and spin that fully characterize a black hole in general
relativity. We show that these images can be significantly different
from the images of a RIAF around a Kerr black hole with the same spin
and demonstrate the feasibility of testing the no-hair theorem by
constraining the quadrupolar deviation from the Kerr metric with
existing EHT data.  Equally important, we find that the disk
inclination and spin orientation angles are robust to the inclusion of
additional parameters, providing confidence in previous estimations
assuming the Kerr metric based upon EHT observations. However, at
present the limits upon potential modifications of the Kerr metric
remain weak.
\end{abstract}


\maketitle

\section{Introduction} \label{sec:I}
Black holes are commonly invoked to power a variety of energetic
phenomena.  Stellar mass black holes are one of the presumed
engines that power gamma-ray bursts \citep{Mesza:06} and lie at the
hearts of X-ray binaries \citep{vdKlis:06}.  Supermassive black holes
are implicated in Active Galactic Nuclei (AGN), powering both
prodigious electromagnetic luminosities and outflows.  In this latter
role supermassive black holes are believed to significantly impact the
structures that surround them on cosmological scales.  This occurs
both by moderating accretion of baryons into dark matter halos
\citep[i.e., feedback; for a review, see][]{Fabi:12} and injecting
material back into the intergalactic medium 
\citep[e.g., magnetic fields;][]{Furl-Loeb:01}.  Thus, supermassive
black holes play a crucial role in coupling the smallest (horizon) and
largest (cosmological) scales in astronomy.

These processes are fundamentally associated with the interaction of
black holes with surrounding matter on scales comparable to the
horizon, via accretion flows, winds, jets, etc.  As such, a
tremendous effort continues to be expended studying the nature of these
astrophysical interactions \citep{Nara:05,Viet:08,Meie:12}.  Critical
to these is the stage set by the black hole, upon which the
astrophysical dynamics unfolds. Thus, imprinted within the
electromagnetic signatures is information regarding the gravitational
properties of the black hole itself. 

Inferring properties of black hole spacetimes through their gravitational
impact upon the dynamics of surrounding matter has been accomplished
successfully in a number of contexts.  Examples include mass
measurements via stellar orbital dynamics in the Galactic center
\citep{Ghez_etal:08,Gill_etal:09b,Gill_etal:09a} and spin measurements
via relativistically broadened Fe K$\alpha$ line measurements
\citep[see, e.g.,][]{Rey:13} and thin disk spectra 
\citep[see, e.g.,][]{McC_Nar_Stei:13}. 
The quasi-periodic oscillations present in the spectra of
some X-ray binaries \citep[see, e.g.,][]{RemMcC:06}, and in some AGN
\citep{GierlQPO:08,ReisQPO:12} present a tantalizing additional
signature, though in the absence of a unique model for their
generation they are presently ambiguous.  In these efforts, general
relativity is accepted as a prior. 

Similar attempts to constrain the applicability of general relativity
have also been made.  These may take the form of ruling out specific
alternative gravity theories,
or by employing one of a growing set of phenomenological
Kerr-like metrics \citep[see, e.g.,][]{PsaltisLRR,Myrz-Seba-Zerb:13}.
These differ from the Kerr metric though they are
not necessarily associated with any particular modified theory of
gravity and, instead, often encompass large classes of such
alternative theories 
\citep[e.g.,][]{MN92,CH04,GB06,VH10,JP11_PRD,Vig11}. 
As with spin, a variety of observations have been brought to bear upon
the applicability of the Kerr metric, including relativistically
broadened iron lines \citep{JP13iron,BambiIron}, variability
\citep{JP11qpo,JP13iron,BambiQPO}, continuum accretion disk spectra
\citep{Bam_Bar:11,Krawc:12,BambiDisk}, and X-ray polarization \citep{Krawc:12}. 
Generally, this is considerably more difficult due to the potential
degeneracy between erroneous accretion physics, which itself is not
fully understood, and deviations from general relativity.

With the advent of millimeter-wavelength very-long baseline
interferometry (mm-VLBI) it has become possible to spatially resolve
horizon-scale structure for a handful of supermassive black holes.
This provides an additional probe of both the gravitational and
accretion/outflow physics.  When combined with preexisting spectral
information, mm-VLBI has proven a strong constraint upon the black
hole properties and provides a potential means to break the
degeneracy between gravitational physics and astrophysics.

The best candidate for horizon-resolving mm-VLBI observations is the
supermassive black hole at the center of the Milky Way, associated
with the bright radio source Sgr A*.  Due to its
proximity, Sgr A* subtends the largest angle on the sky of any known
black hole: with estimates of the mass and distance of
$(4.3\pm0.5)\times10^{6}\,M_\odot$ and $8.3\pm0.4$~kpc, respectively,
the lensed image of Sgr A*'s horizon is 
$53\pm2~{\rm \mu as}$\footnote{The mass and distance measurements are
  strongly correlated, with mass scaling roughly as $M\propto D^{1.8}$,
  resulting in a somewhat smaller uncertainty in the corresponding
  angular size \citep{Ghez_etal:08}.} 
\citep{Ghez_etal:08,Gill_etal:09b,Gill_etal:09a}.
This is well matched to the resolution of existing and future mm-VLBI
experiments, which access angular scales of $\gtrsim30~{\rm \mu as}$
and $\gtrsim10~{\rm \mu as}$, respectively.  In addition, the
spectral energy density of Sgr A* peaks near millimeter wavelengths,
implying that the surrounding accretion flow transitions from being
optically thick to optically thin at these wavelengths.  Hence, the
horizon is likely to be visible.

Unlike AGN, Sgr A* is clearly underluminous, with a
bolometric luminosity of $10^{-9}$ in Eddington units.  This implies
that it is qualitatively different from many AGN, though perhaps more similar
to the remaining 90\% of supermassive black holes that are not
presently active.  While uncertainty remains regarding the morphology
of the emission region, the existing spectral and polarization data
have resulted in a canonical set of components in all models for Sgr A*:
a peaked (often approximated by a Maxwellian) electron distribution
function with a power-law high energy tail and nearly equipartition
magnetic fields.  This results in a wide variety of potential models
for the structure and dynamics of the emission region
\citep[e.g.,][]{Nara_etal:98,Blan-Bege:99,Falc-Mark:00,Ozel-Psal-Nar:00,Yuan-Quat-Nara:03,Loeb-Waxm:07}.  
Nevertheless, the existing mm-VLBI observations have already exhibited
a non-trivial agreement with the RIAF models, suggesting that these
form a credible set of models for the purposes of imaging currently  
\citep{Brod_etal:09,Huan-Taka-Shen:09,Dext-Agol-Frag-McKi:10,Brod_etal:11,Shch-Penn-McKi:12,Dext-Frag:13}.

Here we simulate the first images of a RIAF model around a
supermassive black hole not described by the Kerr metric, and
assess the ability of existing mm-VLBI observations of Sgr~A* to constrain
possible deviations from general relativity.  We do this by exploring
the consequences of the well-developed quasi-Kerr metric, a
parameterized modification of the Kerr metric which contains an
independent quadrupole moment in addition to the mass and spin of the
black hole \citep{GB06}.  This metric provides one of the simplest
frameworks for the study of the effects of a modified quadrupole
moment and parameterizes such deviations in a generic manner. However,
special care must be taken to avoid the region very close to the
central object where the quasi-Kerr metric becomes unphysical
\citep{Johannsen13}.  Our analysis properly takes this requirement
into account and makes a conservative assessment of potential
deviations from the Kerr metric induced by such a quadrupole moment.

Already a variety of potential lensing signatures of deviations from
general relativity within the context of the quasi-Kerr metric have
been identified.  While the shadow of a Kerr black hole is nearly
circular, except for extreme black hole spins and viewing
angles\footnote{Moderate deviations from a circular shadow appear only
  for dimensionless spin parameters of $|a_*|\gtrsim0.9$ oriented at
  large oblique inclinations.}, quasi-Kerr-type quadrupolar metric
deviations typically induce substantial asymmetries \citep{JP10b}.
Similar asymmetries are present for compact objects within the context
of a variety of alternative gravity theories
\citep[see, e.g.,][]{BamYosh10,Amarillaetal10,BamCarMod12,AmaEir12}.

We expand upon these existing studies 
here by repeating the analysis of \citet{Brod_etal:11} employing the
quasi-Kerr metric.  Questions of particular interest include
\begin{itemize}
\item How robust are existing estimates of black hole spin (magnitude
  and direction) to additional metric degrees of freedom?  That is,
  how well is spin measured currently?
\item Which, if any, of the previously identified signatures of
  deviations from the Kerr metric persist once the additional astrophysical
  degrees of freedom are introduced?
\item What constraints can be placed on deviations from the
  Kerr metric with the current EHT data?
\end{itemize}

Generally we find that the black hole spin
orientation is extremely robust to quadrupolar perturbations, with the
inclination and position angle limited to
$\theta={65^\circ}^{+21^\circ}_{-12^\circ}$ and
$\xi={127^\circ}^{+17^\circ}_{-14^\circ}$ (modulo an intrinsic
$180^\circ$ uncertainty), respectively.  Similarly, despite being
strongly correlated with the quadrupolar perturbation, after
marginalization over it, the spin magnitude posterior probability
distribution is also quite robust, implying a dimensionless spin
parameter of $ a_*=0^{+0.7}$.

We find that the images of RIAFs around black holes with nonzero
quadrupolar deviations differ significantly from those without.  We
confirm the general trend of deformed silhouettes in these images.
However, we also identify additional morphological signatures due to
alterations in the dynamical structure of the accretion flow.  These
are not unrelated: at the wavelengths accessible to mm-VLBI the
detailed structure of the black hole shadow (and so-called ``photon
ring'') is partially obscurred by the underlying accretion flow.
Despite this, it remains a key feature in distinguishing the
underlying spacetime metric.

As expected, the current limits upon quadrupolar deviations from mm-VLBI
observations are presently quite weak given the sparse data coverage
of the $u$-$v$ plane.  Nevertheless, features are identified in the
images and mm-VLBI observables that will be accessible to future
observations.

In the remaining subsections we introduce the EHT,
an existing mm-VLBI telescope, and tests of the no-hair theorem using
phenomenological Kerr-like spacetimes.  In Section \ref{sec:QK} we
describe the quasi-Kerr metric, highlighting its features and
limitations.  Section \ref{sec:VM} describes how the accretion flow is
modeled, imaged, and VLBI observables extracted.  Fitting and
parameter estimation procedures are discussed in Section
\ref{sec:VFPE}, with results collected in Section \ref{sec:EIP}.
Finally, conclusions are formulated in Section \ref{sec:C}.
Unless otherwise noted, we adopt units in which $G=c=1$, where $G$ and
$c$ are Newton's constant and the speed of light, respectively.

\subsection{The Event Horizon Telescope}

The EHT is an evolving collection of mm and
sub-mm observatories equipped with VLBI instrumentation for
the purpose of producing horizon-resolving studies of supermassive
black holes at 1.3~mm (230~GHz) and 0.87~mm (345~GHz); see
\citet{Fish_etal:13} for a recent summary of the instrument.
Demonstration experiments were performed in 2007 and 2009, in which
the black hole at the center of the Milky Way, Sgr~A*, was observed 
\citep{Doel_etal:08,Fish_etal:10}.  These provided the first
conclusive evidence for the presence of subhorizon-scale structure in
the emission region of Sgr A*.  These have been followed by
horizon-resolving observations of M87, a known radio jet source
\citep{Doel_etal:12}.

To date, observations by the EHT have been successfully performed with
only three stations: the James Clerk Maxwell Telescope (JCMT) and
Sub-Millimeter Array (SMA) on Mauna Kea in Hawaii, the Combined Array
for Research in Millimeter-wave Astronomy (CARMA) in Cedar Flat,
California, and the Arizona Radio Observatory Sub-Millimeter Telescope
(SMT) on Mount Graham, Arizona.  The longest baseline is roughly
$4.6\times10^{3}$~km, and extends mostly east-west between Hawaii and
the SMT.  Due to the limited number of stations, it is not presently
possible to generate images, requiring that physical analyses be
performed using the visibilities directly. 

Of the two sources observed by the EHT, Sgr A* has been observed most
often and has been the subject of the vast majority of the published
analyses to date.  Thus, here we will focus our attention upon Sgr A*
as well.  The currently available data for Sgr A* is comprised of
observations during two days in 2007 
\citep[April 11 and 12,][]{Doel_etal:08} and three days 
in 2009 \citep[April 5--7,][]{Fish_etal:10}.  Within both observing
periods Sgr A* was in a quiescent state (i.e., did not exhibit any
flaring).  Nevertheless, substantial flux variations were observed,
both between 2007 and 2009 and on day timescales within 2009.  Thus
the data is conveniently separated into four observational epochs
consisting of all days in 2007, and each individual day in 2009.  We
label the former simply as 2007, and the latter by their day, i.e.,
2009.95, 2009.96, and 2009.97.  

In addition to visibility magnitudes, \citet{Fish_etal:10} reported the first
detection of a closure phase on the Hawaii-CARMA-SMT triangle of
$0^\circ\pm40^\circ$.  Despite the large error bars, this is quite
constraining due to the sensitivity of closure phases to the
underlying source structure \citep{Brod_etal:11b}.  Thus, we include
this closure phase measurement in our parameter estimation.

The EHT is expected to grow substantially in the future, both in
sensitivity and in the number of stations.  Additional sites already
in the process of being included are the Atacama Large Millimeter
Array (ALMA) in Chile, the Large Millimeter Telescope (LMT) in Mexico, and
the South Pole Telescope (SPT).  All correspond to long north-south
baselines that critically complement the already existing long
east-west baselines.  Furthermore, the exquisite sensitivity of
ALMA promises to enable the inclusion of a number of smaller sites as
well as dramatic enhancements in the signal-to-noise ratio of existing
sites.  Thus, an assessment of the current and future ability of the
EHT to constrain the properties of the spacetime is especially
timely.  Here we focus upon the current capability, outlining the
procedure and discussing the implications for previous analyses.  We
will treat the future capability elsewhere.

\newpage
\subsection{Parametric Tests of the No-Hair Theorem}

According to the no-hair theorem, a stationary black hole in general
relativity is uniquely characterized by only two parameters\footnote{
In principle, stationary black holes can also have an electric
charge.  However, in astrophysical settings, in which ions are
available in copious quantities, it is expected that gravitationally
relevant charges will neutralize rapidly.}, its mass $M$ and its
spin $J$, and is described by the Kerr metric 
\citep[see, e.g.,][]{Heusler:96}.  Dynamical features, associated,
e.g., with black hole mergers, are expected to damp on timescales
comparable to the light crossing time of the black hole, which even
for the most massive cases is on the order of days.  Hence, it is
generally anticipated that all astrophysical black holes are well
described by the Kerr metric.

Several tests of the no-hair theorem have been proposed to date.
These may be conveniently grouped into tests that probe the spacetime
at locations far from and near to the horizon.  Examples of the former
include high-precision studies of stellar orbital dynamics around
Sgr~A* \citep{Will:08,Merritt:10} and timing measurements of
pulsar-black hole binaries \citep{Wex_Kop:99,Liu_etal:12}, which have
taken on new importance following the discovery of a pulsar in the
Galactic center \citep{Kenn_etal:13}. 
Proposed tests of the strong-field properties of astrophysical black
holes are either based on the gravitational waves generated by stellar-mass
compact objects falling into supermassive black holes in extreme
mass-ratio inspirals \citep[EMRIs; for a review, see][]{Gair_LRR} or
on the electromagnetic radiation emitted from black hole accretion
disks \citep{JP10a}.  In order to properly model the expected signals
of these tests, both approaches generally modify the Kerr metric using
a parametrically deformed Kerr-like spacetime, which contains one or
more additional free parameters and which reduces to the Kerr metric
if all deviations vanish 
\citep[e.g.,][]{MN92,CH04,GB06,VH10,JP11_PRD,Vig11}. A
measurement of these deviations can then provide a parameterized test
of the applicability of the Kerr metric to astrophysical black holes
\citep{Ryan:95}.  

Here we employ a multipole expansion of the Kerr metric.
In analogy to Newtonian gravity, where the potential $\Phi$ of a given
mass distribution can be expanded in a series of spherical harmonics
whose coefficients are the multipole moments of the source mass, in
general relativity a multipole expansion can be constructed that
describes the mass and matter current distribution of 
the spacetime \citep[see, e.g.,][]{Thorne:80}.
The no-hair theorem immediately implies that all higher order
multipole moments of a Kerr black hole are not independent, but rather
can be expressed in terms of the mass and spin.  This relationship can
be expressed particularly compactly via 
\begin{equation}
M_l + i S_l =  M (i a)^l
\label{eq:Kerrmult}
\end{equation}
where $M_l$ and $S_l$ are the mass and matter current multipole
moment, and $a\equiv J/M$ where $J$ is the black hole angular
momentum \citep{Geroch:70,Hansen:74}.  In this expansion, mass and
angular momentum may be identified with the first two multipoles 
of the spacetime, i.e., $M_0=M$ and $S_1= J$.  Of particular
importance here is the next order, the quadrupole mass moment:
\begin{equation}
Q_{\rm K} \equiv M_2 = -Ma^2\,.
\label{eq:QKerr}
\end{equation}

One of the simplest ways to introduce a deviation from the Kerr metric
is to construct a quasi-Kerr metric \citep{GB06} as the underlying 
spacetime, for which the quadrupole may be set independently from the
spin, having the form 
\begin{equation}
Q_{\rm QK} = -M(a^2 + \epsilon M^2),
\label{eq:QQKerr}
\end{equation}
where $\epsilon$ is a dimensionless parameter that measures
potential deviations from the Kerr metric.  When $\epsilon=0$, the quasi-Kerr
metric reduces to the Kerr metric, and equation~(\ref{eq:QKerr}) is trivially
recovered.  

Modifying the spacetime of a black hole in this manner
leads to important changes of its properties that may be used to
empirically test the no-hair theorem \citep{JP10a}. 
Most important for our purposes here are the modifications of the null
geodesic structure of the spacetime.  This results in characteristic
distortions of the black hole shadow, the lensed image of the black
hole horizon cast against a putative surrounding accretion flow, and
the surrounding photon ring, which corresponds to 
the projection along null geodesics of the orbits of photons that
propagate around the black hole many times before they are
observed. While this ring is exactly circular for a Schwarzschild
black hole and nearly circular for a Kerr black hole with spin values
$ a/M\lesssim0.9$, the shape of this ring becomes asymmetric if the
no-hair theorem is violated \citep{JP10b}.  It is the prospect of
detecting these signatures that we are assessing here.

For convenience, we also define a dimensionless spin parameter
$a_*\equiv a/M$, often referred to simply as ``spin'' in the literature,
and the gravitational radius $\Rg\equiv GM/c^2$, in terms of which the
Schwarzschild radius is simply $2\Rg$.

\section{The Quasi-Kerr Metric} \label{sec:QK}

\begin{figure*}[ht!]
\begin{center}
\includegraphics[width=0.3\textwidth]{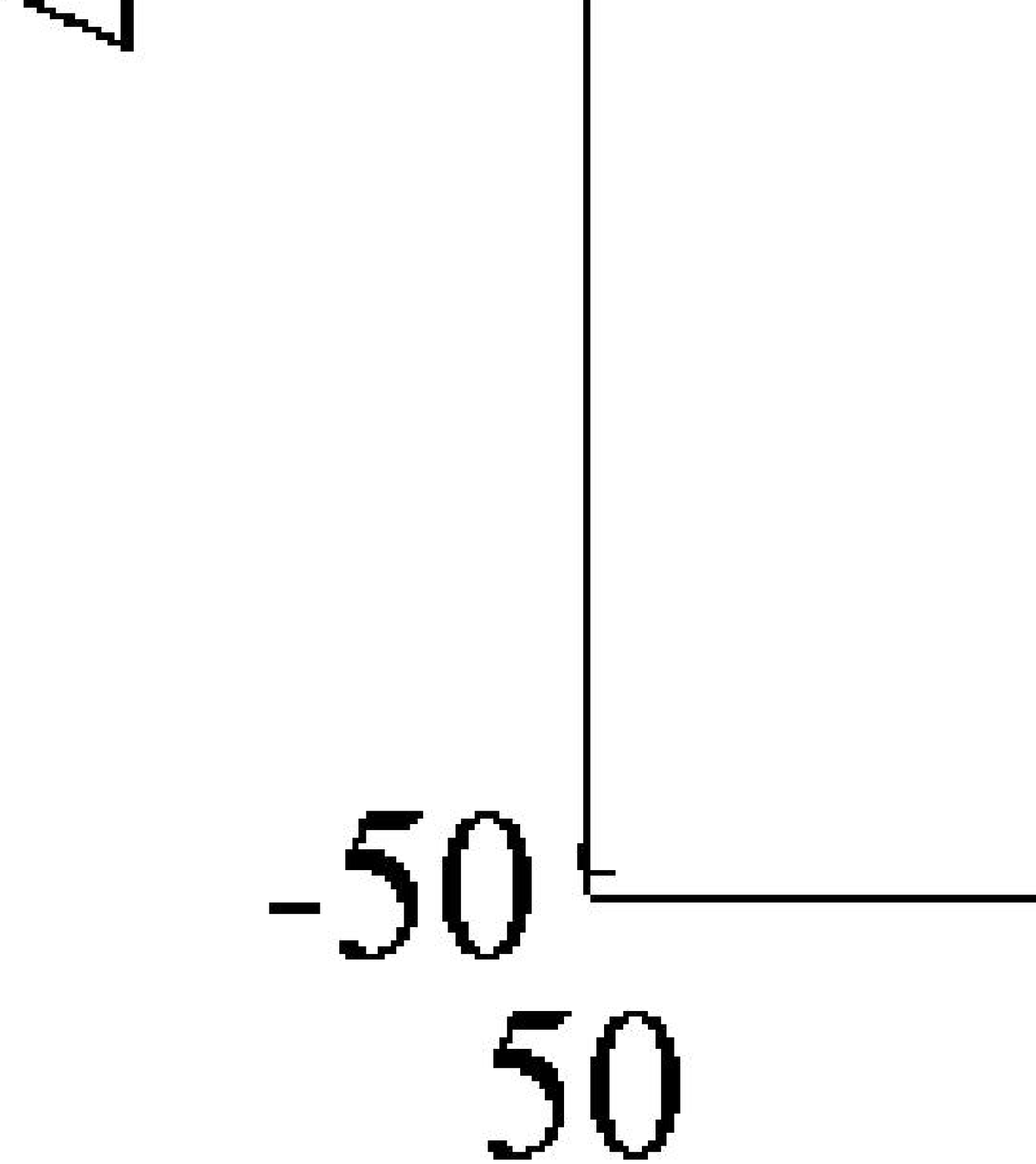}
\includegraphics[width=0.3\textwidth]{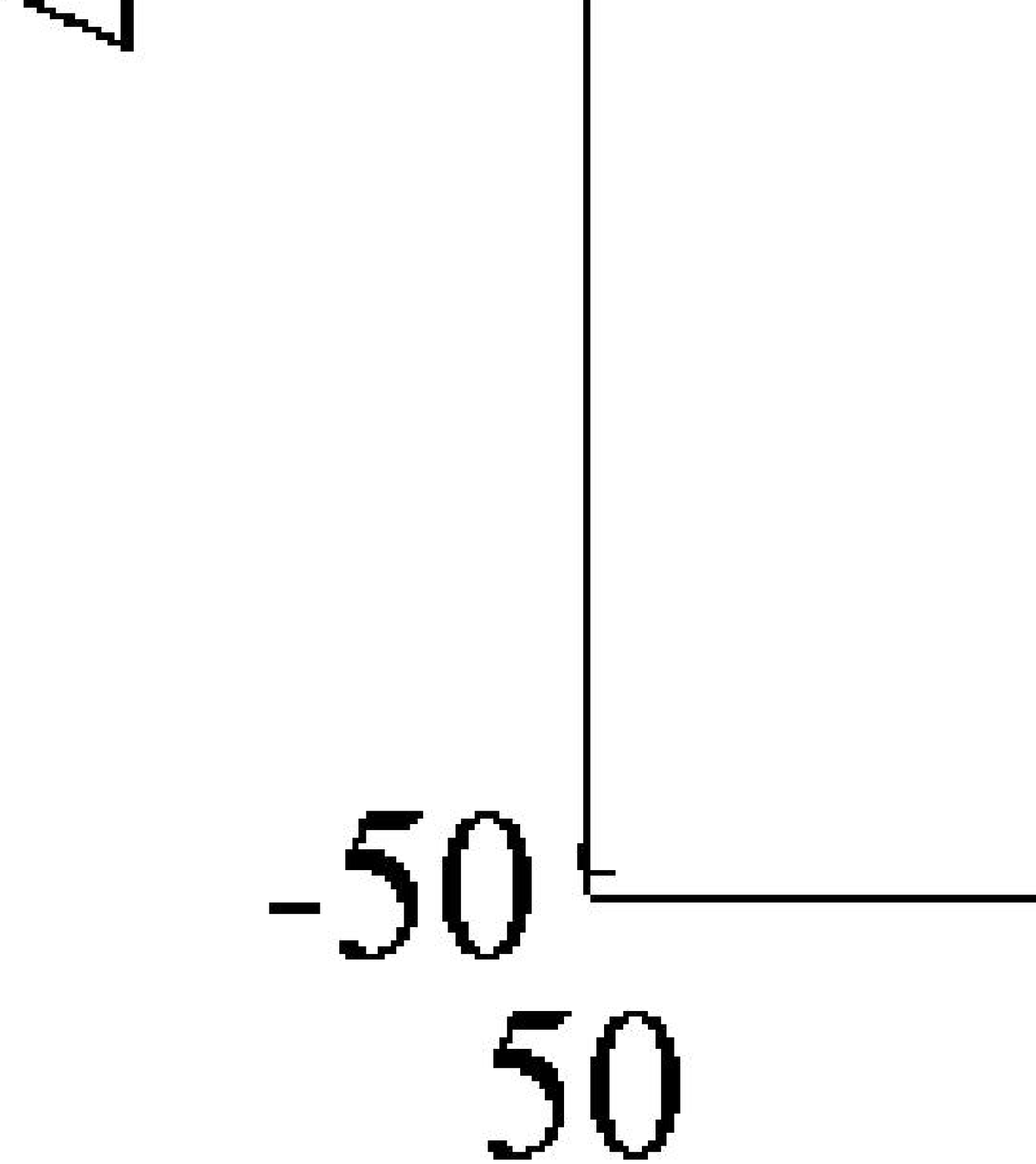}
\includegraphics[width=0.3\textwidth]{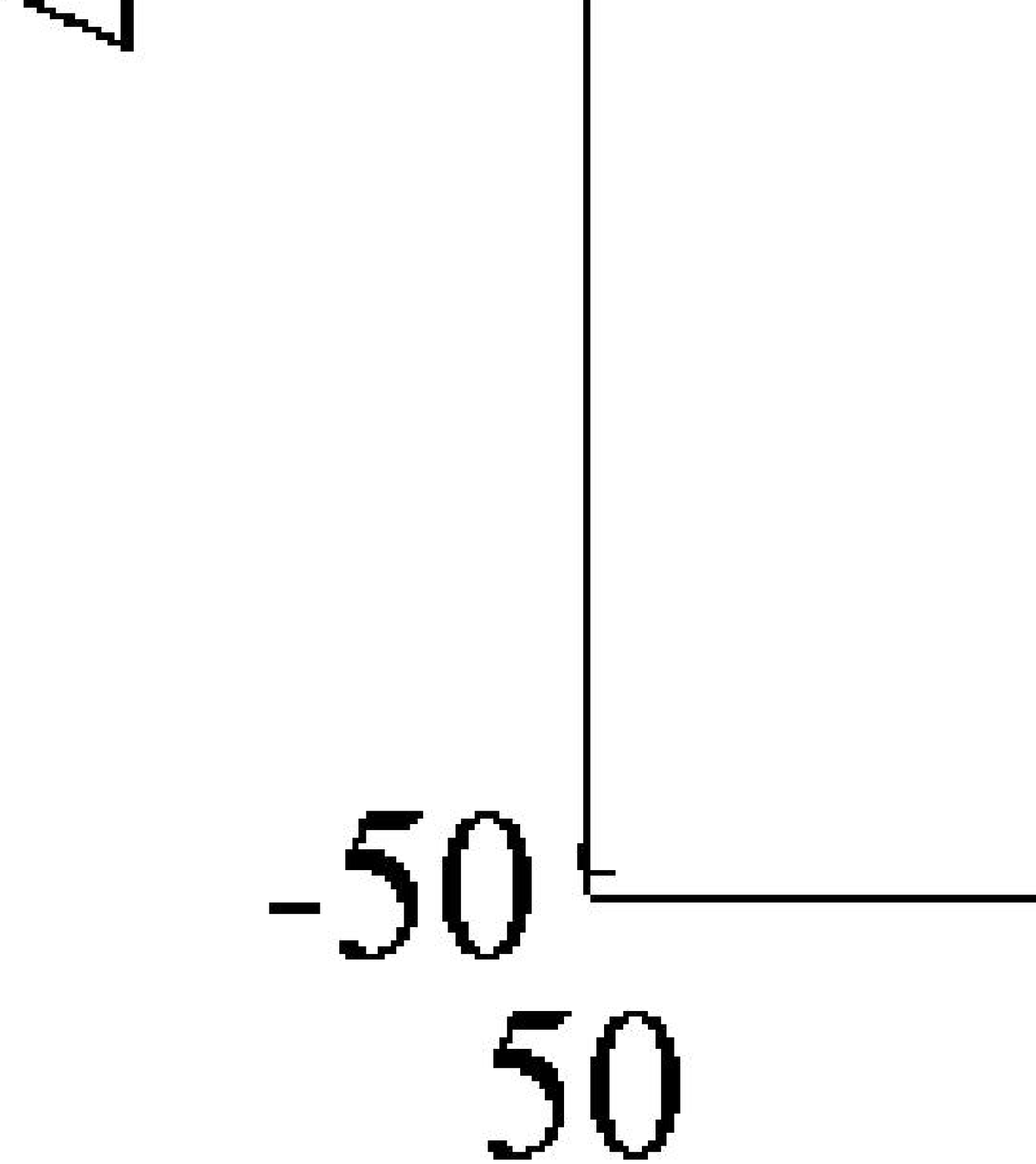}
\end{center}
\caption{Example black hole silhouettes for non-rotating black holes
  with illustrative values of the quadrupolar parameter, and applying
  a cutoff radius of $r_{\rm cutoff}=3\Rg$, as described in the
  text.  In all cases $\theta=90^\circ$ and the coordinate axis is
  parallel to the vertical direction. For negative values of the
  deviation parameter $\epsilon$ (left panel), the shadow has a more
  prolate shape than the shadow of a Kerr black hole with the same
  spin (central panel), while for positive values of the deviation
  parameter (right panel), the shadow has a more oblate
  shape.}\label{fig:sil} 
\end{figure*}

In this section, we describe the quasi-Kerr metric \citep{GB06} and
summarize some of its properties, which were analyzed in detail in
\citet{JP10a} and \citet{Johannsen13}. The quasi-Kerr metric
\citep{GB06} derives from the Hartle-Thorne metric
\citep{Hartle:67,HT:68}, originally designed to describe the exterior
spacetimes of slowly rotating neutron stars.  \citet{GB06} considered
this metric as a small perturbation of the Kerr metric and used it to
construct approximate waveforms for EMRIs around a non-Kerr compact
object.  In this context, the quasi-Kerr spacetime becomes accessible
to particles at very small radii that otherwise lie below the neutron
star surface and is formally applicable to black holes with small
spins.  This metric, then, generally harbors a 
naked singularity\footnote{This may be explicitly verified by a
  careful examination of the Kretschmann scalar, 
  $K\equiv R^{\alpha\beta\gamma\delta} R_{\alpha\beta\gamma\delta}$,
  where $R^{\alpha\beta\gamma\delta}$ is the Riemann curvature
  tensor.} \citep{Johannsen13}.  Subsequently, \citet{JP10a}
appropriated this metric as a framework for tests of the no-hair
theorem based on the electromagnetic radiation emitted from the
accretion flows around black holes.

Treating the quasi-Kerr metric as a small perturbation to the Kerr
metric, however, is not entirely possible within the context of
ray-tracing computations, because the geodesic equations are nonlinear
and the perturbative nature of the metric cannot be consistently
enforced.  Instead, we treat the quasi-Kerr metric as an ``exact''
metric with a deformed quadrupole moment and with the same
higher-order multipole moments as the Kerr metric,
permitting the study of the effects of a deviation from the Kerr
metric at the quadrupole order \citep[see the discussion in][]{Johannsen13}.  
Therefore, no longer constrained by the need to maintain a
perturbation expansion, and given that such deviations are largely
unconstrained empirically, we do not limit our analysis to small
values of spin either, considering values up to $ a_*=0.9$.
However, precautions must be taken to avoid any contact of
photons and particles in the accretion flow with the naked 
sigularity at small radii, which restricts the range of quadrupolar
deviations that we may consider such that the quasi-Kerr metric
remains a valid framework for our analysis \citep[see][]{Johannsen13}.
We describe these restrictions in detail below.

In Boyer-Lindquist coordinates $(t,r,\theta,\phi)$, the quasi-Kerr
metric can be described by the line element \citep{GB06} 
\begin{equation}
ds^2 = g_{tt}dt^2 + 2g_{t\phi}dtd\phi + g_{rr}dr^2 + g_{\theta\theta}d\theta^2 + g_{\phi\phi}d\phi^2,
\end{equation}
where
\begin{equation}
\begin{aligned}
g_{tt} 
&= -\left(1-\frac{2Mr}{\Sigma}\right) + \frac{5\epsilon(1+3\cos2\theta)}{32M^2r^2}\\
&\quad\times \bigg[ 2M \left( 3r^3-9Mr^2+4M^2r+2M^3 \right)\\
&\qquad\qquad - 3r^2(r-2M)^2 \ln\left( \frac{r}{r-2M} \right) \bigg]\,,\\
g_{t\phi} 
&= -\frac{2Mar\sin^2\theta}{\Sigma}\,,\\
g_{rr} 
&= \frac{\Sigma}{\Delta} - \frac{5\epsilon(1-3\cos^2\theta)}{16M^2(r-2M)^2} \\
&\quad\times \bigg[ 2M(r-M)\left(3r^2-6Mr-2M^2\right) \\
&\qquad\qquad -3r^2(r-2M)^2 \ln\left( \frac{r}{r-2M} \right) \bigg]\,,\\
g_{\theta \theta} 
&= \Sigma - \frac{5\epsilon r (1+3\cos2\theta)}{32M^2} \bigg[ -2M \left(3r^2+3Mr-2M^2\right) \\
&\qquad + 3r\left(r^2-2M^2\right) \ln\left( \frac{r}{r-2M} \right) \bigg]\,,\\
g_{\phi \phi} 
&= \bigg[r^2+a^2+\frac{2Ma^2r\sin^2\theta}{\Sigma} - \frac{5\epsilon r (1+3\cos2\theta)}{32M^2} \\
&\quad\times \bigg[ -2M \left(3r^2+3Mr-2M^2\right) \\
&\qquad\qquad + 3r\left(r^2-2M^2\right) \ln\left( \frac{r}{r-2M} \right) \bigg] \bigg]\sin^2\theta\,,
\end{aligned}
\label{QKmetric}
\end{equation}
in which
\begin{equation}
\Delta \equiv r^2-2Mr+a^2
\quad\text{and}\quad
\Sigma \equiv r^2+a^2\cos^2 \theta\,.
\label{deltasigma}
\end{equation}

For nonzero values of the parameter $\epsilon$, the properties of
the quasi-Kerr spacetime are altered in several key aspects \citep{JP10a}: (i)
the locations of the circular photon orbit and the
innermost stable circular orbit (ISCO) are shifted; (ii) trajectories
of photons near the compact object are gravitationally lensed to a
stronger or lesser degree; and (iii) observed photon redshifts
are altered, both potentially increased or decreased. All of these
features lead to a modified image of any putative accretion flow.

Very close to the naked singularity, the quasi-Kerr metric contains
regions where the metric signature is no longer Lorentzian and where
closed timelike curves exist \citep{Johannsen13}. Since all of these
pathologies are unphysical, we introduce a cutoff radius, 
$r_{\rm cutoff}$, outside of which the metric is well behaved.  Any
particle that enters the region where $r<r_{\rm cutoff}$ is considered
captured by the black hole.  Thus, the cutoff radius acts as an
artificial horizon, which, for the purposes of this paper, allows us
to effectively treat the central object as a black hole and ensures
that all unphysical regions are shielded from distant observers
\citep{JP10a}.

Here we choose to place the cutoff at a fixed radius 
$r_{\rm cutoff}=3 \Rg$, the radius of the circular photon orbit of
a Schwarzschild black hole, which facilitates the comparison of images
with different values of the spin and the deviation parameter.
With this definition of the cutoff radius we ignore the emission and
propagation of photons inside this region.  This implies that for some
values of $a$ and $\epsilon$ the photon orbit lies within 
$r_{\rm cutoff}$, and therefore the innermost portion of the image has
been possibly neglected.  As might be expected, this results in a
marginally larger black hole shadow.  More importantly, the neglected
part of the image includes the photon ring, the shape of which
directly imprints the structure of the spacetime \citep{JP10b}.  Thus,
by imposing a cutoff radius we necessarily ignore features that are
particularly sensitive to quadrupolar deviations for
$\epsilon\lesssim0$.  Extending $r_{\rm cutoff}$ inside of $3 \Rg$
would generally result in more extreme deviations, and thus stronger
limits upon $\epsilon$. 

Furthermore, to avoid possible artifacts associated with our choice of
$r_{\rm cutoff}$, we restrict our attention to parameter choices for
which the ISCO is larger than $4 \Rg$.  Since the location of the ISCO
is a function of both the spin and the deviation parameter $\epsilon$, 
this necessarily limits the parameter space we may consider, with the
minimum permissible value of the parameter $\epsilon$ being a function
of the spin.  At vanishing spin, permissible metrics can be found at for
$\epsilon\gtrsim-0.8$, below which the ISCO lies within $4 \Rg$.
Thus, for concreteness, we restrict ourselves to
$-0.8\le\epsilon\le1$.    Since high spin has already been found to
be highly disfavored, we also limit ourselves to $0\leq a_*\leq0.9$.
However, the condition upon the ISCO implies that not all
$(a,\epsilon)$ within these ranges are permissible: higher spin
values generally result in smaller ISCOs, and therefore more stringent
limits upon the deviation parameter (see Figure~\ref{fig:pmulti}).   

In practice, $r_{\rm cutoff}$ represents a choice for how solutions to
a modified theory of strong gravity will alter the Kerr metric
near the horizon.  Without the introduction of a cutoff radius, images
of accretion flows would reveal what we consider to be artificial
features that arise from the pathological regions that lie inside of
this radius.  In particular, for positive values of the parameter
$\epsilon$, images show two bright spots, which are located along the
spin axis of the black hole inside of the domain of the shadow. For
negative values of the parameter $\epsilon$, images contain a
ring-like band that stretches across the center of the image.  These
features arise due to regions in which $g_{rr}<0$, i.e., the compact
object has effectively become repulsive.  By excluding these regions
from our analysis we necessarily restrict ourselves to observable
signatures arising from regions in which the quasi-Kerr metric is well
defined.  Hence, in this sense, we are making the most conservative
comparison possible with this context.

The shadow cast by the black hole is shown in Figure~\ref{fig:sil} for
a non-rotating black hole at various values of $\epsilon$.  The
qualitative impact is precisely that expected by \citet{JP10a}: 
more prolate spacetimes ($\epsilon<0$) produce extended silhouettes
along the coordinate axis, while more oblate spacetimes
($\epsilon>0$) produce extended silhouettes in the equatorial
plane.  Even for order unity quadrupolar deviations the distortions
are relatively small in comparison to the unperturbed shadow, 
increasing the semi-major axis of the shadow by roughly 7\% at
$|\epsilon|\simeq0.8$.  For Sgr A*, this implies variations in the
image structure on scales of order $4$ $\mu$as or smaller.
Nevertheless, we will see that this degree of image variations is
readily distinguishable even with the existing mm-VLBI data.

\section{Visibility Modeling} \label{sec:VM}

Due to the limited $u$-$v$ coverage of the early EHT observations of
Sgr A* it is not presently possible to construct high-fidelity image
reconstructions from the reported visibilities.  Nor is it clear that
analyzing images is intrinsically more desirable.  Thus it is necessary
to construct visibility models based on physical models of the
spacetime and the surrounding emission regions.  In the previous
section, we described the spacetime, from which it is straightforward
to compute photon trajectories (i.e., null geodesics).  Here we
describe the model we use for the emission region, based upon the RIAF
models employed in \citet{Brod_etal:11}.  Because this has been
described at length there (and references therein) here we provide
only a cursory summary of the model details, referring the interested
reader to those for the particular justifications.

Thus we expect that the limits upon $\epsilon$ obtained
here are pessimistic in practice. Note that for this reason
the parameter estimates from the existing EHT data at $\epsilon=0$
are slightly different from those found in \citet{Brod_etal:11}.

\subsection{Accretion Flow Modeling} \label{sec:VM:AFM}

Sgr A* transitions from an inverted, presumably optically thick
spectrum to an optically thin spectrum near millimeter wavelengths.
This implies that near 1.3~mm Sgr A* is only becoming optically thin,
and thus absorption in the surrounding medium is likely to be
important.  This transition does not occur isotropically, happening at
longer wavelengths for gas that is receding and at shorter wavelengths
for gas that is approaching.  Therefore, proper modeling the
structure and relativistic radiative transfer is crucial to producing
high fidelity images.

Metric perturbations impact the accretion flow in a number of ways,
substantially modifying its dynamics and the lensing of the emitted
radiation, as well as potentially affecting its spatial structure.
To directly compare with the analyses in \citet{Brod_etal:11}, we will
ignore alterations to the flow structure (in a coordinate sense), and
focus solely upon the observational signatures associated with lensing
and dynamics.  This is motivated in part by the dominance of the
latter effects observed in studies of the impact of black hole spin,
and suggested below (Section \ref{sec:Imps}) where it is found that
spacetime modifications are manifested in the resulting images
primarily via the location of the ISCO (albeit within the constrained
model class described here).  It is also a result of expedience; a
full exploration of self-consistent disk modeling in modified
spacetimes is beyond the scope of this paper.

For concreteness, as in \citet{Brod_etal:11}, we follow
\citet{Yuan-Quat-Nara:03} and employ a model in which the accretion
flow has a population of thermal electrons with density and temperature
\begin{equation}
n_{e,{\rm th}}=n_{e,{\rm th}}^0 \left(\frac{r}{2 \Rg}\right)^{-1.1} e^{-z^2/2\rho^2}
\end{equation}
and
\begin{equation}
T_{e}=n_{e}^0 \left(\frac{r}{2 \Rg}\right)^{-0.84}\,,
\end{equation}
respectively,
and a toroidal magnetic field in approximate ($\beta=10$)
equipartition with the ions (which are responsible for the majority of
the pressure), i.e.,
\begin{equation}
\frac{B^2}{8\pi} = \beta^{-1} n_{e,{\rm th}} \frac{m_p c^2 \Rg}{6 r}\,.
\end{equation}
In all of these, $\rho$ is the cylindrical radius and $z$ is the
vertical coordinate.

Beyond the ISCO the accretion flow orbits
with the angular velocity of the equatorial stable circular orbit with
the same cylindrical radius, i.e., the relativistic analogs of
Keplerian orbits\footnote{With the exception that the quasi-Kerr
  metric is employed, this is identical to the orbital velocities
  consider in \citet{Brod_etal:11}.}.  Inside of the ISCO we assume
the gas is plunging upon ballistic trajectories.  In principle the
plunging gas can still radiate, though in practice it contributes
little to the overall emission due to the large radial velocities it
develops.

In the case of the thermal quantities the radial structure
was taken from \citet{Yuan-Quat-Nara:03}, and the vertical structure
was determined by assuming that the disk height is comparable to
$\rho$.  Note that all of the models we employ necessarily have the
spin aligned with the orbital angular momentum of the accretion flow.
For the regions that dominate the $\mm$ emission it is unclear if this
is a good assumption for RIAF models generally and Sgr A*
specifically.  Assessing the restriction this implies will be
considered in future work.

Thermal electrons alone are incapable of reproducing the nearly-flat
spectrum of Sgr A* below $43\,\GHz$.  Thus it is necessary to also
include a nonthermal component.  As with the thermal components, we
adopt a self-similar model for a population of nonthermal electrons,
\begin{equation}
n_{e,{\rm nth}}=n_{e,{\rm nth}}^0 \left(\frac{r}{2\Rg}\right)^{-2.02} e^{-z^2/2\rho^2}\,,
\end{equation}
with a power-law distribution corresponding to a spectral index of
$1.25$ and cut off below Lorentz factors of $10^2$ 
\citep[consistent with][]{Yuan-Quat-Nara:03}.  The radial power-law
index was chosen to reproduce the low frequency spectrum of Sgr A*,
and is insensitive to the black hole properties due to the distant
origin of the long-wavelength emission.

The primary emission mechanism at the wavelengths of interest is
synchrotron radiation, arising from both the thermal and nonthermal electrons.
We model the emission from the thermal electrons using the emissivity
described in \citet{Yuan-Quat-Nara:03}, appropriately altered to
account for relativistic effects \citep[see, e.g., ][]{Brod-Blan:04}.
Since we perform polarized radiative transfer via the entire
complement of Stokes parameters, we employ the polarization fraction
for thermal synchrotron as derived in \citet{Petr-McTi:83}.  In doing
so, we have implicitly assumed that the emission due to thermal
electrons is isotropic, which while generally not the case is unlikely
to change our results significantly.  For the nonthermal electrons, we
follow \citet{Jone-ODel:77} for a power-law electron distribution,
with an additional spectral break associated with the minimum electron
Lorentz factor.  For both emission components the absorption
coefficients are determined directly via Kirchhoff's law.  Images are
then produced using the fully relativistic ray-tracing and radiative
transfer schemes described in \citet{Brod-Loeb:06a,Brod-Loeb:06b} and
\citet{Brod:06}, into which the quasi-Kerr metric described in Section
\ref{sec:QK} is inserted.

Because \citet{Yuan-Quat-Nara:03} neglected relativistic effects and
assumed spherical symmetry, their models are not directly applicable
here.  For these reasons, as in \citet{Brod_etal:11}, at each point of
interest in the $a$-$\theta$-$\epsilon$ parameter space, the
coefficients $(n_{e,{\rm th}}^0,T_e^0,n_{e,{\rm nth}}^0)$ were
adjusted to fit the radio spectral energy distribution (SED) of Sgr A* 
\citep[for details on the fitting procedure, see][]{Brod_etal:11}.
In all cases it was possible to fit the SED with extraordinary precision.

Due to its time-consuming nature, the spectral fitting procedure was
performed at a subset of points in the $a$-$\theta$-$\epsilon$
parameter space.  Explicitly, for values of the spin and the deviation
parameter for which the ISCO exceeded the $4\Rg$ limit, fits were 
performed at values of $\epsilon$ ranging from $-0.8$ to $1$ in steps
of $0.1$, $a$ from $0$ to $0.9$ in steps of $0.1$, and $\theta$ from
$0^\circ$ to $90^\circ$ in steps of $10^\circ$.
Values for the fitted coefficients 
$(n_{e,{\rm th}}^0,T_e^0,n_{e,{\rm nth}}^0)$ where then obtained
at arbitrary points via high-order polynomial interpolation.

During the $1.3\mm$-VLBI observations, Sgr A*'s flux varied by roughly
$30\%$.  We model this as a variable accretion rate, allowing the
electron density normalization to vary, corresponding to a modulation
of the accretion rate.  In practice, we reduced
the electron density normalization by an amount sufficient to produce
a total flux of $2.5\,\Jy$, and then multiplied the resulting images
by a correction factor during the EHT data analysis.  Because the
source is not uniformly optically thin, this is not strictly correct,
though the associated error in the image structure is small.

Finally, arbitrary position angles, $\xi$, are modeled simply by 
rotating the images on the sky.  For this purpose we define $\xi$ as
the position angle (east of north) of the projected spin vector.

\begin{figure*}
\begin{center}
\includegraphics[width=0.3\textwidth]{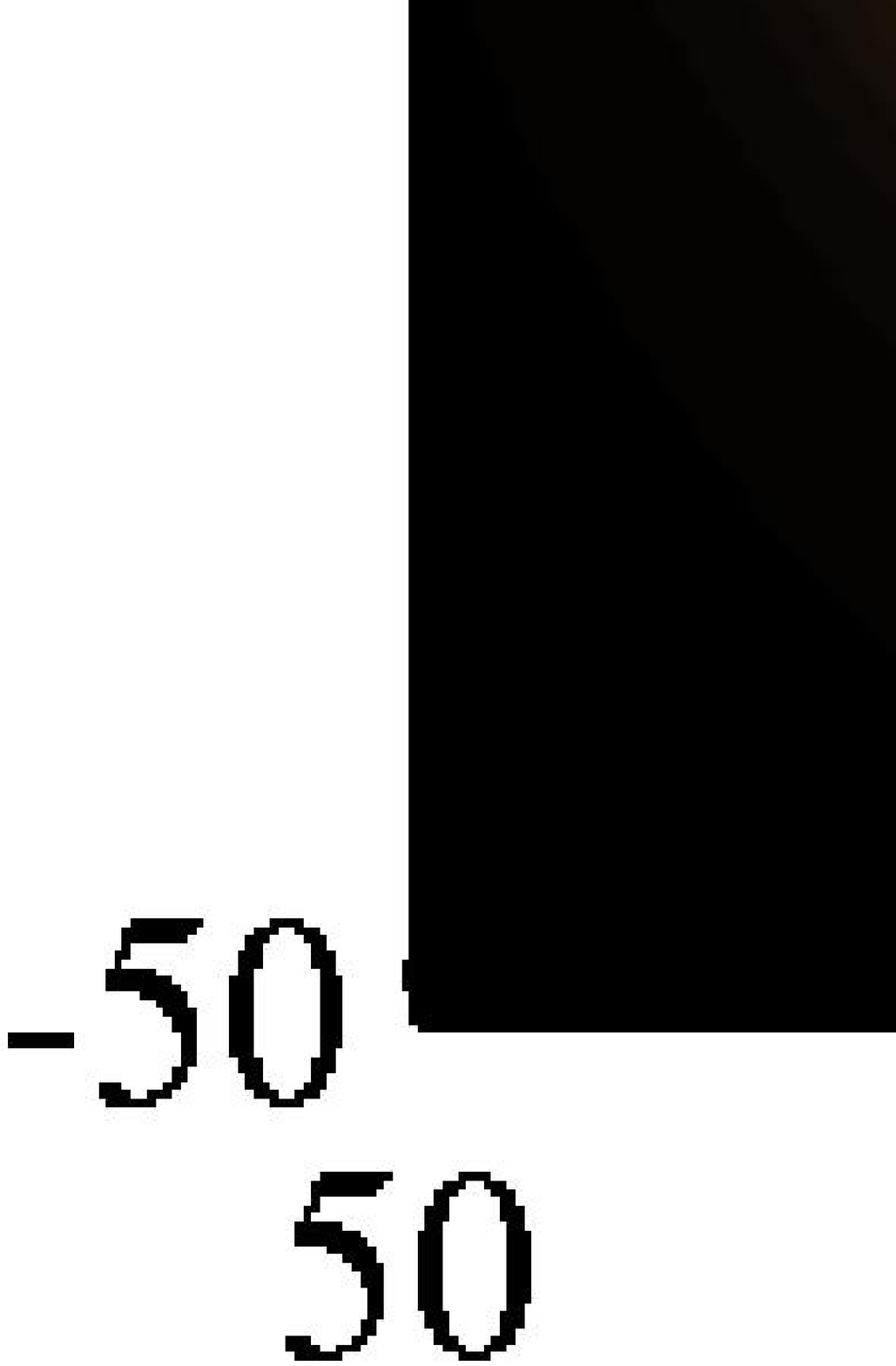}
\includegraphics[width=0.3\textwidth]{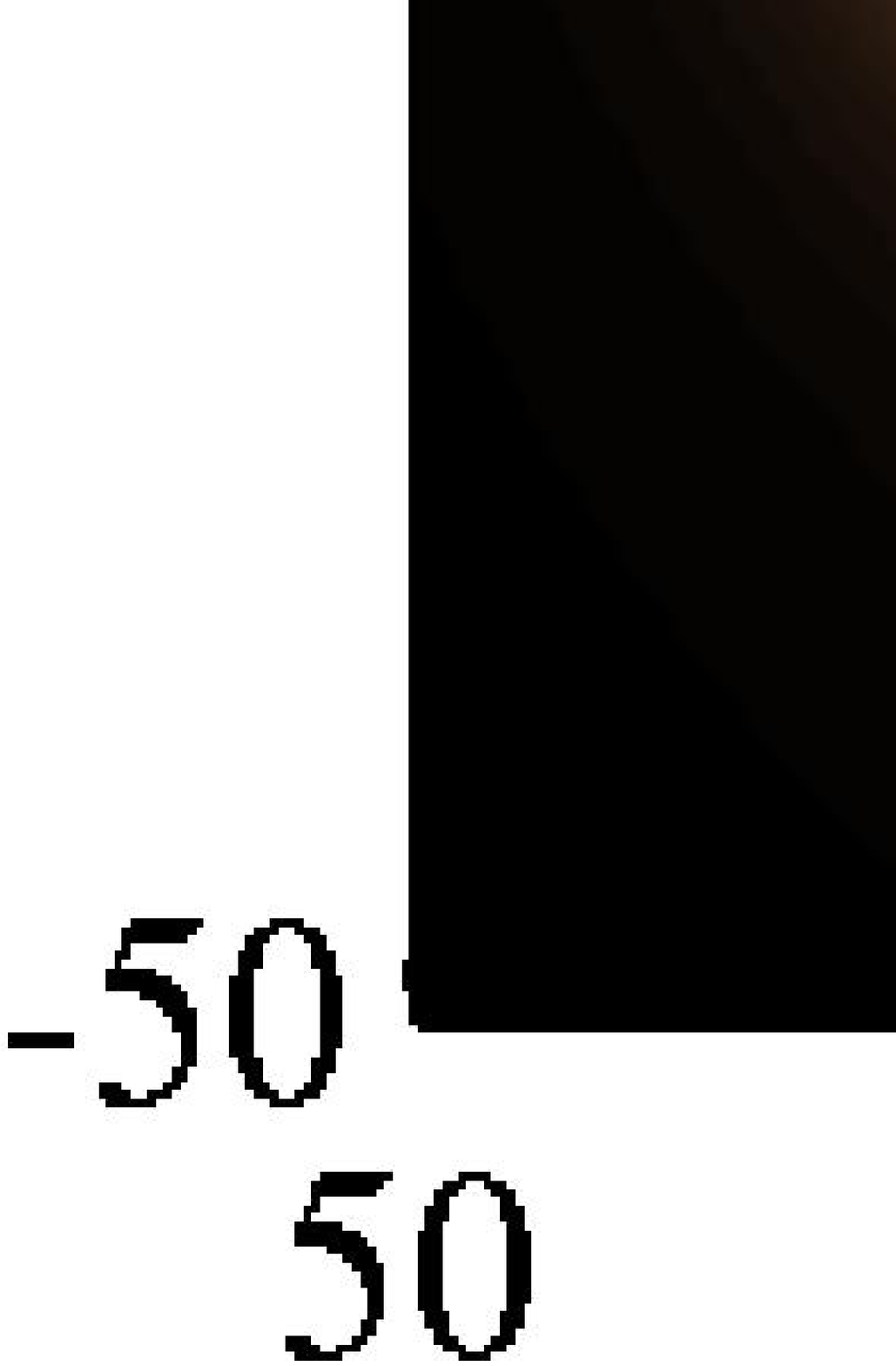}
\includegraphics[width=0.3\textwidth]{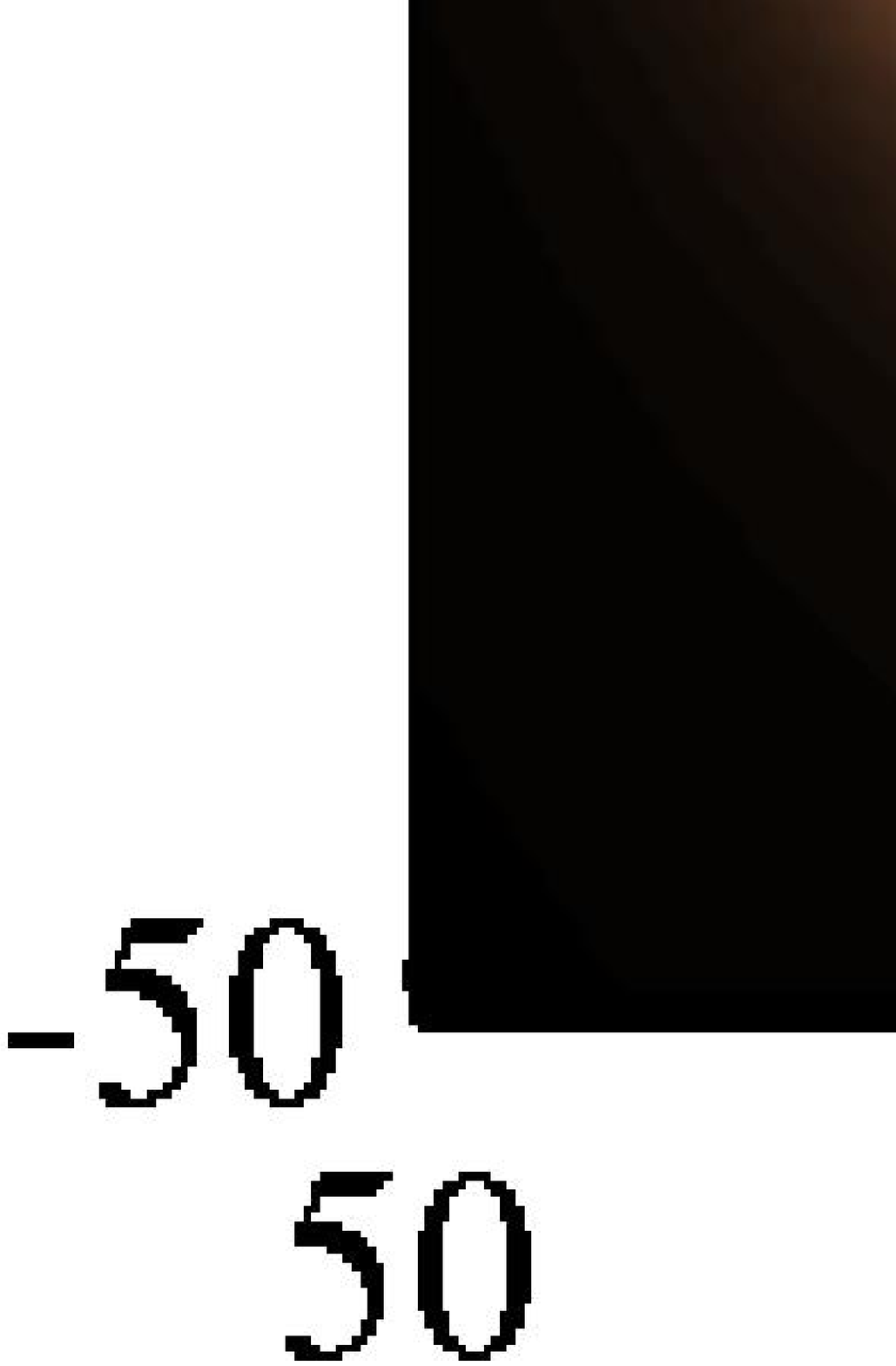}
\end{center}
\caption{Spectrally fit radiatively inefficient accretion flow model
  images at 230~GHz, constructed as described in Section
  \ref{sec:VM:AFM}, with $ a_*=0$ and $\theta=90^\circ$ and
  $\xi=0^\circ$ for the same values of the quadrupolar deviation
  parameter $\epsilon$ shown in Figure \ref{fig:sil}.  In all cases,
  the shadow silhouette is clearly visible even though it is partially
  obscured by the accretion flow on the left side of the shadow which
  approaches the observer. Nonzero values of the deviation parameter
  $\epsilon$ modify the morphology and measured intensity of the
  crescent and, for sufficiently negative values of the deviation
  parameter, the crescent acquires a more pronounced tongue-like flux
  feature in the equatorial plane of the black
  hole.}\label{fig:images} 
\end{figure*}

A collection of example images constructed via the above procedure is
presented in Figure \ref{fig:images}, corresponding to the same
spacetimes used to construct the silhouettes in Figure \ref{fig:sil}.
In all cases they exhibit the crescent morphology characteristic of
thick accretion flows, arising due to the relativistic orbital
motions: both Doppler beaming and boosting result in an enhancement
and reduction of the surface brightness and optical depth on the
approaching (left) and receding side (right) of the accretion flow,
respectively.  

In all images the silhouette of the black hole is clearly visible,
varying qualitatively as anticipated by Figure \ref{fig:sil}.
The detailed structure of the shadow, however, is partially obscured by the
optically thick crescent on the approaching side of the accretion
flow.  Thus, as expected, the additional degrees of freedom in the
accretion flow do slightly moderate the impact of the spacetime upon the
apparent shape of the shadow itself (cf. Figures \ref{fig:images} and
\ref{fig:sil}).

However, more importantly, additional features not present in the
silhouettes alone appear.  These include the equatorial tongue-like flux
feature in the left panel of Figure \ref{fig:images}, 
which is primarily caused by the enhanced Doppler boosting of the
accretion flow particles near the ISCO, and the variation 
in the crescent width, due to modifications in the orbital dynamics of
test particles.  Thus, the images contain a substantial amount of
information about the spacetime, both directly via the shadow shape
and the null geodesic structure and indirectly via the impact of the
spacetime upon the matter dynamics and hence the emission and
subsequent radiative transfer through the accretion flow.

\subsection{Computing Visibility Magnitudes and Closure Phases}
Given a trial image intensity distribution, $I(\alpha,\beta)$, where
$\alpha$ and $\beta$ are angular coordinates, we may compute the
visibilities in the standard fashion: 
\begin{equation}
V(u,v) = \int\int \d\alpha \d\beta\, e^{-2\pi i(\alpha u + \beta v)/\lambda} I(\alpha,\beta)\,.
\end{equation}
To compute the visibility magnitudes, these must then be modified to
account for the blurring effects due to electron scattering resulting
from propagation through the Galactic plane.

The effect of interstellar electron scattering in the direction of
Sgr A* has been carefully characterized empirically by a number of
authors.  This has been found to be consistent with convolving the
source with an asymmetric Gaussian, with major axis nearly aligned
with east-west, and a $\lambda^2$ wavelength dependence.  We employ
the model from \citet{Bowe_etal:06}, which has major axis oriented
$78^\circ$ east of north, with associated full width at half-maximum
for the major and minor axes given by
\begin{equation}
\begin{aligned}
{\rm FWHM}^{\rm ES}_M &= 1.309\left(\frac{\lambda}{1\,\cm}\right)^{2}\,\mas\,,\\
{\rm FWHM}^{\rm ES}_m &= 0.64\left(\frac{\lambda}{1\,\cm}\right)^{2}\,\mas\,,
\end{aligned}
\end{equation}
respectively.  In practice, the interstellar electron scattering
convolution was effected in the $u$-$v$ plane, where the convolution
reduces to a multiplication.

Finally, closure phases are obtained for a given triplet of VLBI
stations via their definition: the closure phase for a given triplet of
observatories is then given by
\begin{equation}
\Phi_{ijk}
=
\arg\left[V(u_{ij}, v_{ij})\right]
+\arg\left[V(u_{jk}, v_{jk})\right]
+\arg\big[V(u_{ki}, v_{ki})\big]\,.
\end{equation}
In practice, where this required some interpolation, this was
performed upon the real and imaginary components of $V$ and then used
to reconstruct the phases.

\section{Visibility Fitting and Parameter Estimation} \label{sec:VFPE}

Comparing the models described in Section \ref{sec:VM} with the
existing EHT observations presents both practical and conceptual
difficulties.  The construction of large libraries of spectrally-fit
models is a computationally expensive exercise, which becomes
exponentially harder as additional parameters are considered.
Here we describe the scheme by which our model library was constructed
and summarize the statistical measures of the fits employed.

\subsection{Model Library Construction and Comparison Efficiency}
The analysis of the EHT data within the context of RIAF models
which assume general relativity described in \citet{Brod_etal:11}
already taxed the substantial computational resources available.  This
was a result both of the process of spectral fitting as well as the
large-scale visibility generation and comparisons.  Thus, the addition
of a quadrupole parameter required the development of new analysis
techniques.  These came in two forms: increasing the speed with which
a model at a given point in the parameter space can be compared to the
data, and more efficiently choosing regions within the parameter space
to search for satisfactory models.

\subsubsection{Improved Model Comparison Efficiency}
In practice, comparing a model to the EHT data is performed via
the following steps.
\begin{enumerate}
\item Compute an image of the model with the given parameters at
  $1.3\,\mm$ with an appropriate flux normalization.
\item From the image, compute the complex visibilities via an FFT, up
  to an uncertain epoch-dependent flux normalization.
\item Obtain the complex visibilities at the points in the $u$-$v$
  plane at which the visibility magnitudes were measured.  Note that
  in practice this is done for a large number of position angles at a
  time, corresponding to a rotation of the image.
\item From the complex visibilities compute the magnitudes at the
  relevant $u$-$v$ positions, and closure phases on the relevant
  triangles.
\item Construct a likelihood by direct comparison to the data.
\item Marginalize or maximize over the epoch-dependent flux normalizations.
\end{enumerate}
The first step requires large-scale computing to complete in a
reasonable amount of time, typically weeks.  Optimizing the first is
subject of the following subsection.  The remaining steps are
performed on a workstation, typically taking days.  This is dominated by
the time associated with constructing the likelihoods and
marginalizing over the flux normalizations.  The reason is the
presence of a visibility magnitude upper-limit in 2007, which requires
computing an error-function and, most importantly, forcing a numerical
approach to the marginalization.  

However, it is possible to analytically perform the marginalizations
if only detections (and more specifically, measurements with Gaussian
errors) are considered (see Appendices \ref{app:Fnorm} and
\ref{app:MLP}).  Repeating the analysis in \citet{Brod_etal:11} at a lower 
parameter space resolution, we have verified that in light of the
detections in the 2009 epochs the 2007 CARMA-Hawaii upper-limit is no
longer significantly constraining.  By proceeding without the
upper-limit the analysis is sped up by roughly an order of magnitude.

A second benefit of restricting our attention to detections, is that
the maximum and marginalized flux normalizations are the same.  Thus,
it is only necessary to compute one set of quantities to address both
the posterior probability density of a given set of parameters (the
marginal value) and assess the quality of the fit via a $\chi^2$
and/or Bayesian analysis (the maximum value).

\begin{figure*}
\begin{center}
\includegraphics[width=0.45\textwidth]{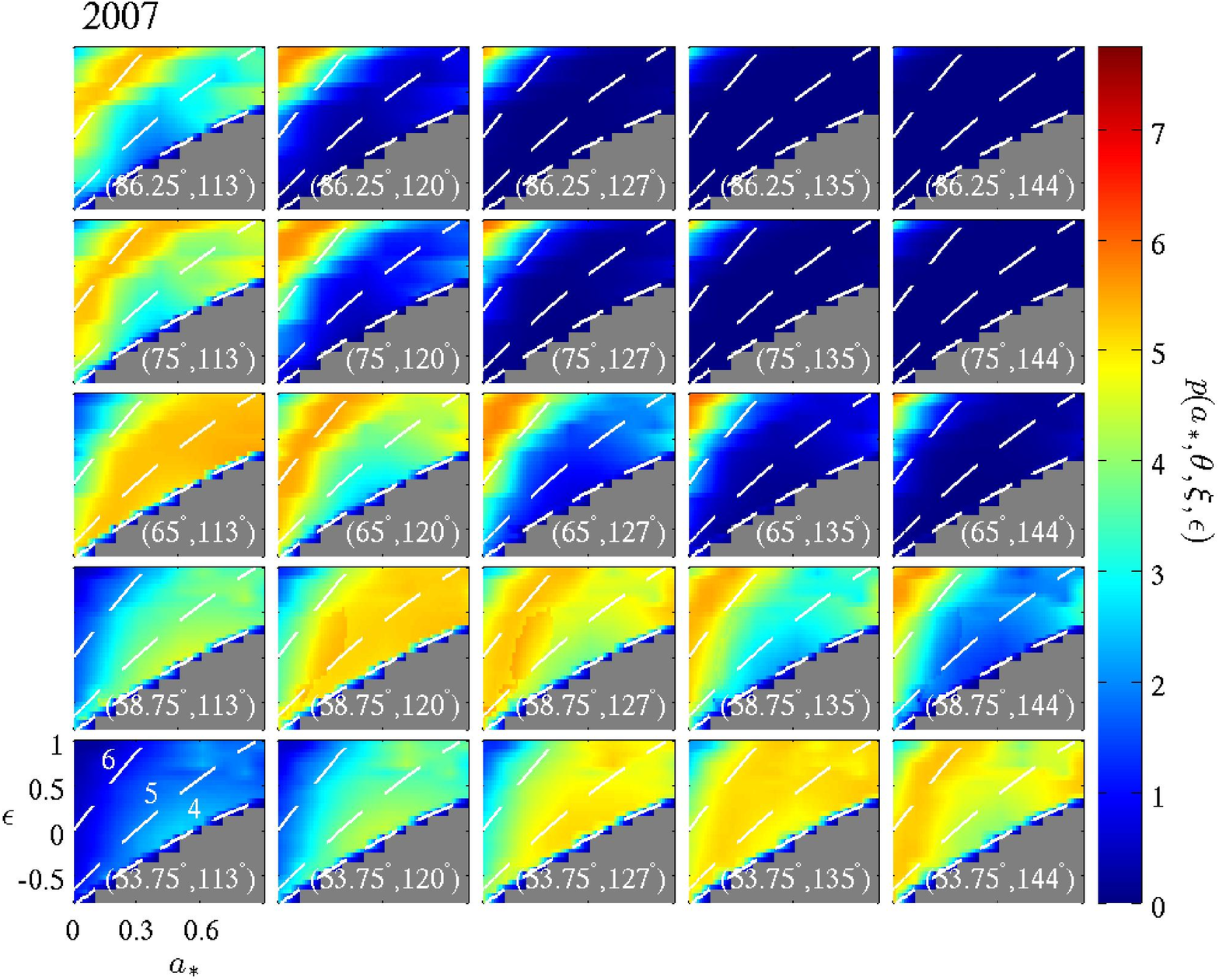}
\includegraphics[width=0.45\textwidth]{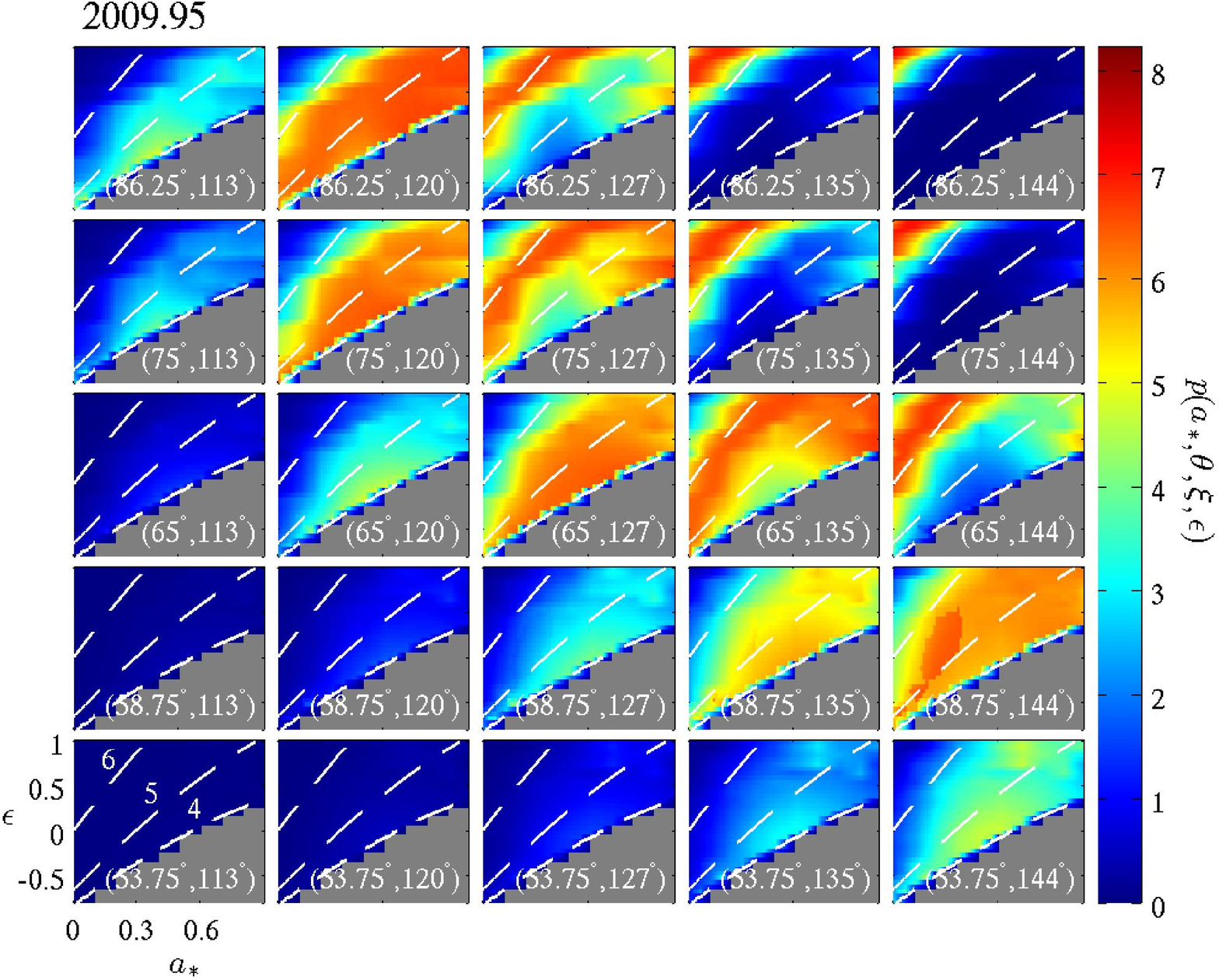}\\
\includegraphics[width=0.45\textwidth]{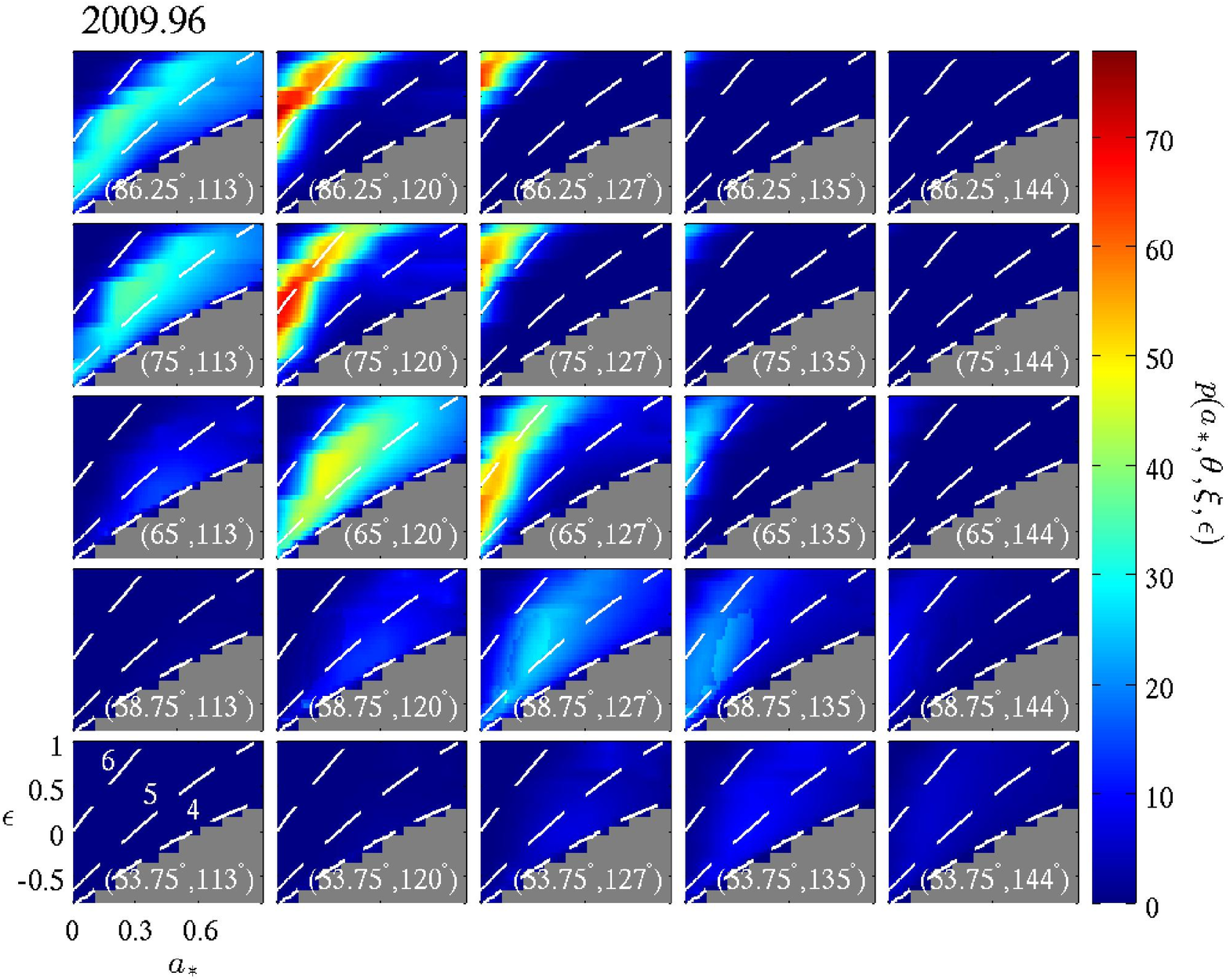}
\includegraphics[width=0.45\textwidth]{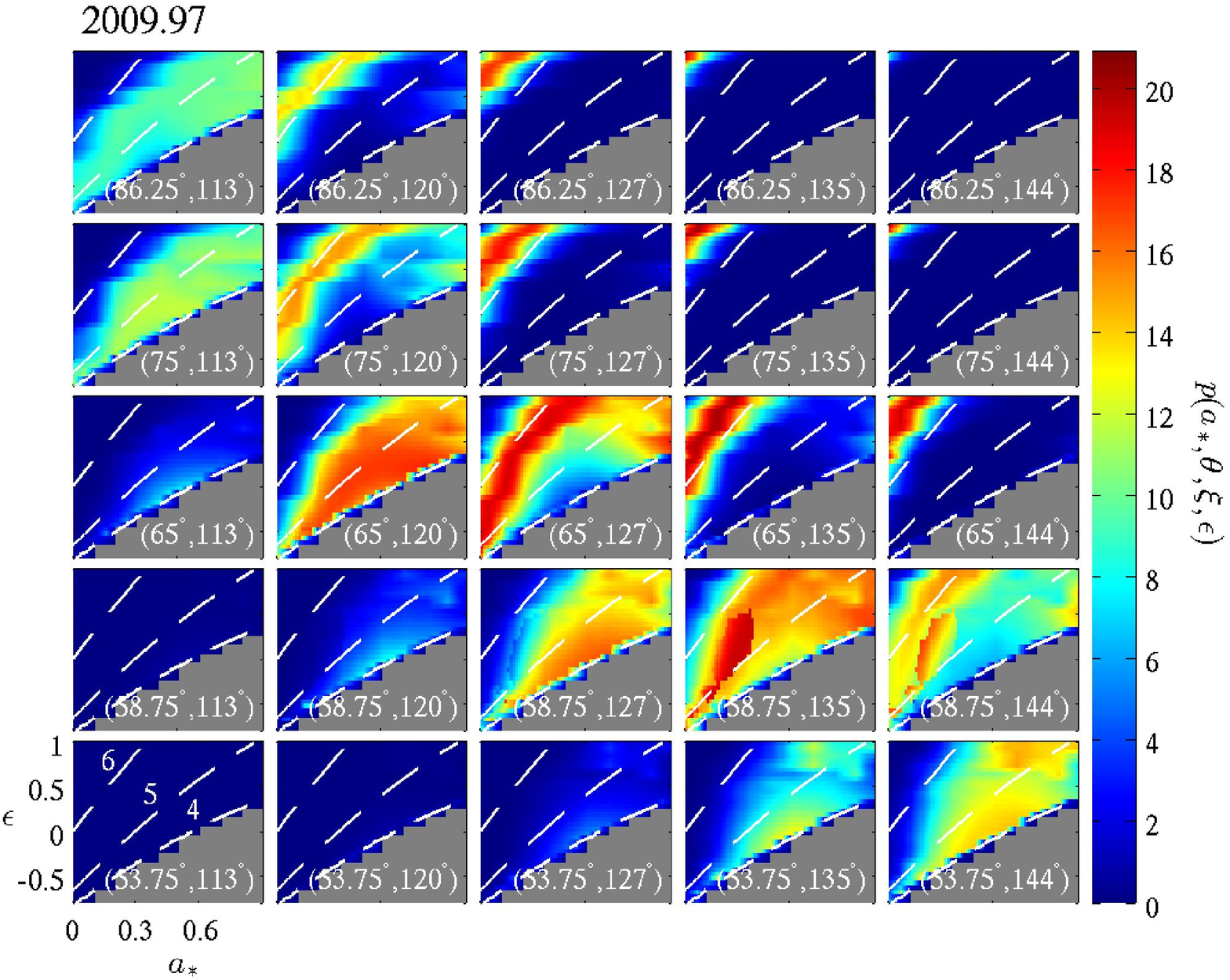}
\end{center}
\caption{Posterior parameter probability distributions
  $p(a,\theta,\xi,\epsilon)$ for each observation epoch.  Each
  multipanel plot shows $p(a,\theta,\xi,\epsilon)$, normalized by the
  average probability density, as a function of $a$ and $\epsilon$ for
  a number of $(\theta,\xi)$ pairs.  The range in angles bound the
  1$\sigma$ region when marginalized over $a$ and $\epsilon$.  The
  grayed region denotes parameter values for which the ISCO lies
  within $4\Rg$, and were therefore excluded from consideration.  For
  reference, lines of constant ISCO are shown by the white dashed
  contours, corresponding to $6\Rg$, $5\Rg$, and $4\Rg$ (top
  to bottom).} \label{fig:pmulti}
\end{figure*}

\subsubsection{Adaptive Parameter Space Refinement}
The \citet{Brod_etal:11} analysis was performed with an image library
consisting of nearly 10,000 images, and once the model library was
constructed (by far the most time consuming step), took roughly a
month to complete.  Even an only moderate sampling in the new
parameter, $\epsilon$, would result in an intractable problem.  Due to
the current data quality, it is still necessary to probe large
portions of the possible parameter space, with the ultimately
acceptable models constituting a strongly correlated, clearly
non-Gaussian subset.

The solution we have adopted is to perform repeated lower-resolution
analyses, zooming in on the high-probability region.  That is, the
first image sample, comprised of roughly 1,000 images, is analyzed to
obtain a rough measure of the likelihood throughout the parameter
space.  All regions with likelihoods larger than 10\% of the maximum
value were then selected for refinement, where the resolution was
doubled, resulting in another 3,000 images.  In regions with
likelihoods greater than 50\% of the new maximum, a third set of
images, roughly 2,500, were produced and subsequently analyzed.
Finally, a third round of refinement is performed where the new
likelihoods are again above 50\% of the maximum value, resulting in
9,000 new images.

At each refinement, the probabilities and likelihoods are interpolated
to a parameter-space grid of twice the resolution, with values in the
refinement regions being obtained by the new analyses.  Thus, this
produces an inhomogeneous coverage of the parameter space volume,
focusing attention on high-probability regions but nevertheless
probing the whole space at least at low resolution.  At the highest
resolution we have $\Delta a_*=0.01252$,
$\Delta\theta=1.25^\circ$, $\Delta\xi=1^\circ$, and
$\Delta\epsilon=0.05$.  Apart from the introduction of the parameter
$\epsilon$, these are marginally coarser than, though comparable
to, those employed in \citet{Brod_etal:11}.

\section{Estimation and Interpretation of Parameters} \label{sec:EIP}

The primary product of the forgoing analysis is the construction of a
set of likelihoods at each point in the 4-dimensional black hole
parameter space ($a$, $\theta$, $\xi$, $\epsilon$), marginalized over
the four epoch-specific flux normalizations,
$\bar{L}(a,\theta,\xi,\epsilon)$.  From this we can
construct the posterior probability distributions of the parameters in
the normal way (see Appendix \ref{app:MLP}).  In doing so we adopt
flat priors on $a$ and $\epsilon$ (though see Section
\ref{sec:Imps}), and an isotropic prior on $(\theta,\,\xi)$.
With the likelihoods and posterior parameter probabilities, here we
assess the quality of the fits, obtain estimates for the black hole
parameters and uncertainties, and discuss their implications.

\begin{figure*}
\begin{center}
\includegraphics[width=\textwidth]{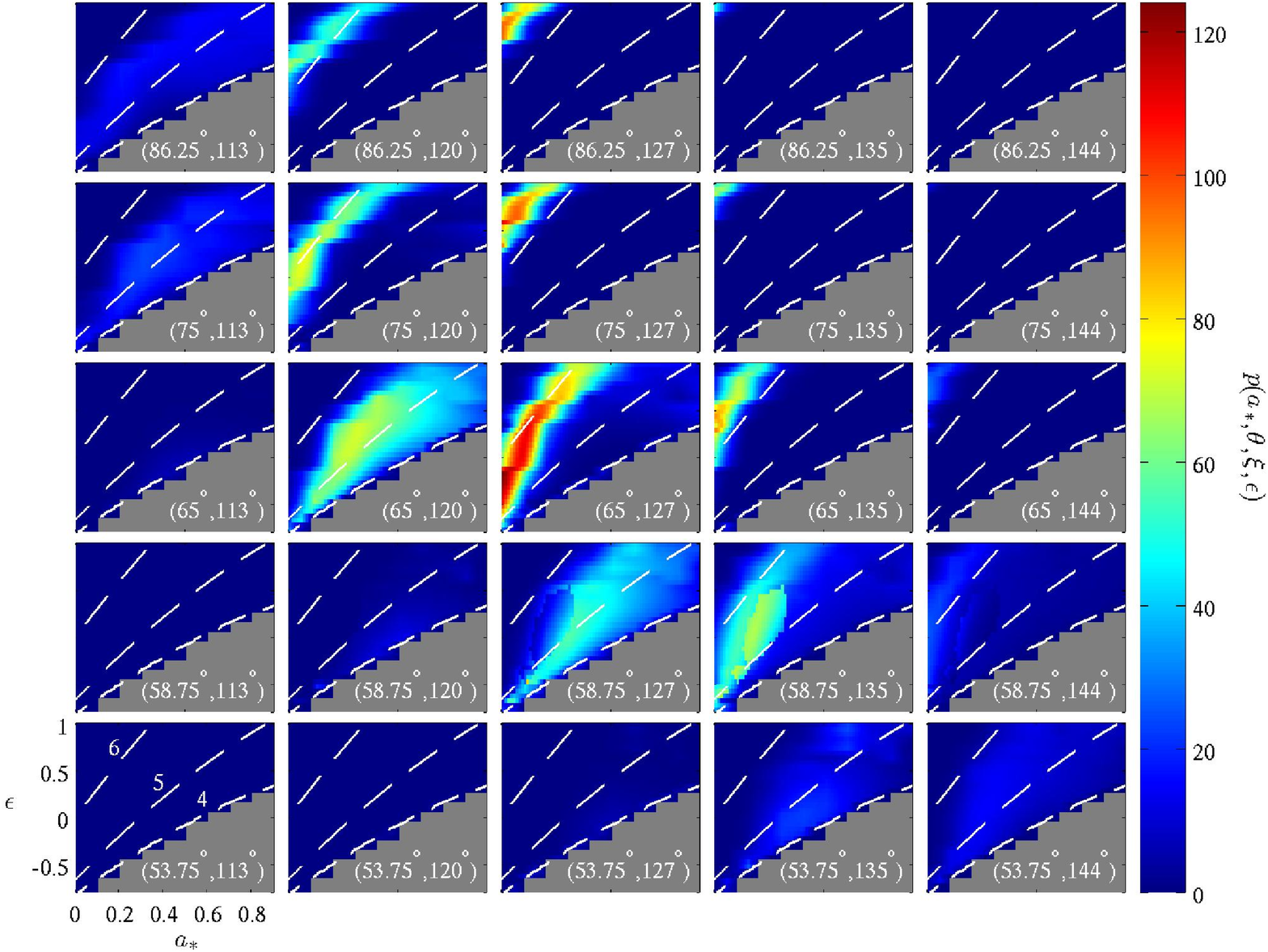}
\end{center}
\caption{Combined posterior parameter probability distribution for all
  observation epochs, $p(a,\theta,\xi,\epsilon)$, normalized by the
  average probability density, as a function of $a$ and $\epsilon$
  for a number of $(\theta,\xi)$ pairs.  The range in angles bound the
  1$\sigma$ region when marginalized over $a$ and $\epsilon$.  The
  grayed region denotes parameter values for which the ISCO lies
  within $4\Rg$, and were therefore excluded from consideration.
  For reference, lines of constant ISCO are shown by the white dashed
  contours, corresponding to $6\Rg$, $5\Rg$, and $4\Rg$ (top
  to bottom).} \label{fig:pall}
\end{figure*}

\subsection{Assessing Fit Quality}

\begin{deluxetable}{lcccccc}\tabletypesize{\small}
\tablecaption{Model Fitting Results Summary\label{tab:stats}}
\tablehead{
\colhead{Model} &
\colhead{$k$\tablenotemark{a}} &
\colhead{$\chi^2$} &
\colhead{${\rm DoF}$\tablenotemark{b}} &
\colhead{$\chi^2/{\rm DoF}$} &
\colhead{$\BIC$\tablenotemark{c}} &
\colhead{$\AIC$\tablenotemark{d}}
}
\startdata
Broderick et al. (2011) & 7 & 53.09 & 64 & 0.830 & 82.9 & 68.9\\
$\epsilon=0$   & 7 & 51.79 & 65 & 0.797 & 81.7 & 67.5\\
All $\epsilon$ & 8 & 51.74 & 64 & 0.809 & 86.0 & 70.0
\enddata
\tablenotetext{a}{Number of model parameters, consisting of epoch flux
  normalizations and assumed black hole parameters.}
\tablenotetext{b}{Associated degrees of freedom; the data used here
  includes the closure phase reported in \citet{Fish_etal:10}, and
  thus includes one more point than used in \citet{Brod_etal:11}.}
\tablenotetext{c}{Bayesian information criterion, for a full
  definition see \citet{Brod_etal:11}.}
\tablenotetext{d}{Akaki information criterion, for a full
  definition see \citet{Brod_etal:11}.}
\end{deluxetable}

Prior to parameter estimation and interpretation, we address the
existence of ``good'' fits, i.e., high likelihood fits.
Given the results of \citet{Brod_etal:11}, we might anticipate the
presence of good fits is guaranteed.  Indeed, despite the small
variation due to the choice to excise the photons passing within
$3\Rg$ and the inclusion of the closure phase measurement in the
analysis, we find that this remains the case.  Collected in Table
\ref{tab:stats} are the log-likelihood ($\chi^2$), number of model
parameters ($k$), and various measures of fit quality for the analysis
in \citet{Brod_etal:11}, the analysis here restricted to $\epsilon=0$
and in full.  These measures include the Bayesian and Akaki
information criteria as defined in \citet{Brod_etal:11}.

Despite the inclusion of the closure phase, the distinction between
the analysis in \citet{Brod_etal:11} and here when restricted to Kerr is 
insignificant, as expected.  Allowing $\epsilon\ne0$ improves the
log-likelihood only marginally at the expense of an additional
parameter.  Thus, as will be verified explicitly in Section
\ref{sec:Lims} below, little evidence exists within the existing
EHT data to support the need for a non-Kerr quadrupole.

Since the black hole parameters are expected to remain fixed over the
current duration of the EHT observations, we also assess the
consistency of the fits across epochs.  The epoch specific posterior
probability densities are shown in Figure \ref{fig:pmulti} (note the 
differing color scales for each epoch).  In these plots artifacts of
the adaptive refinement scheme are clearly visible, though they
primarily affect regions with probability densities below 50\% of the
maximum in each plot.

While at low probability density the epochs exhibit significant
deviations from each other, the epochs are consistent where the
probability densities are highest.  This is most clearly seen by
comparing the remaining epochs to 2009.96 (and, to a lesser degree,
with 2009.97).  All other epochs have substantial posterior probability
densities in regions that are highly probable in 2009.96.  Thus, a
single spacetime structure is consistent with the entire set of EHT
observations.

The aggregate posterior probability densities are shown in Figure
\ref{fig:pall}.  Many of the features exhibited on the individual
epoch analyses (especially 2009.96) persist.  The highest probability
densities form a narrow tongue, aligned most prominently with lines of
constant ISCO (shown by the white dashed lines).  While the
particular ISCO value varies with spin orientation, for the
orientation containing the highest probability density this is at
roughly $6 \Rg$, corresponding to the ISCO of a slightly modified
Schwarzschild black hole with a value of the quadrupolar deviation
parameter $\epsilon=-0.2^{+1.2}_{-0.6}$ and values of the inclination 
$\theta={65^\circ}^{+25^\circ}_{-12^\circ}$ and position angle
$\xi={127^\circ}^{+17^\circ}_{-14^\circ}$.
  Quoted are the bounds upon the
1$\sigma$ envelope within the full four-dimensional parameter space,
though these are potentially misleading since some of the parameters
are significantly correlated (see the following sections).

\begin{figure*}
\begin{center}
\includegraphics[width=\textwidth]{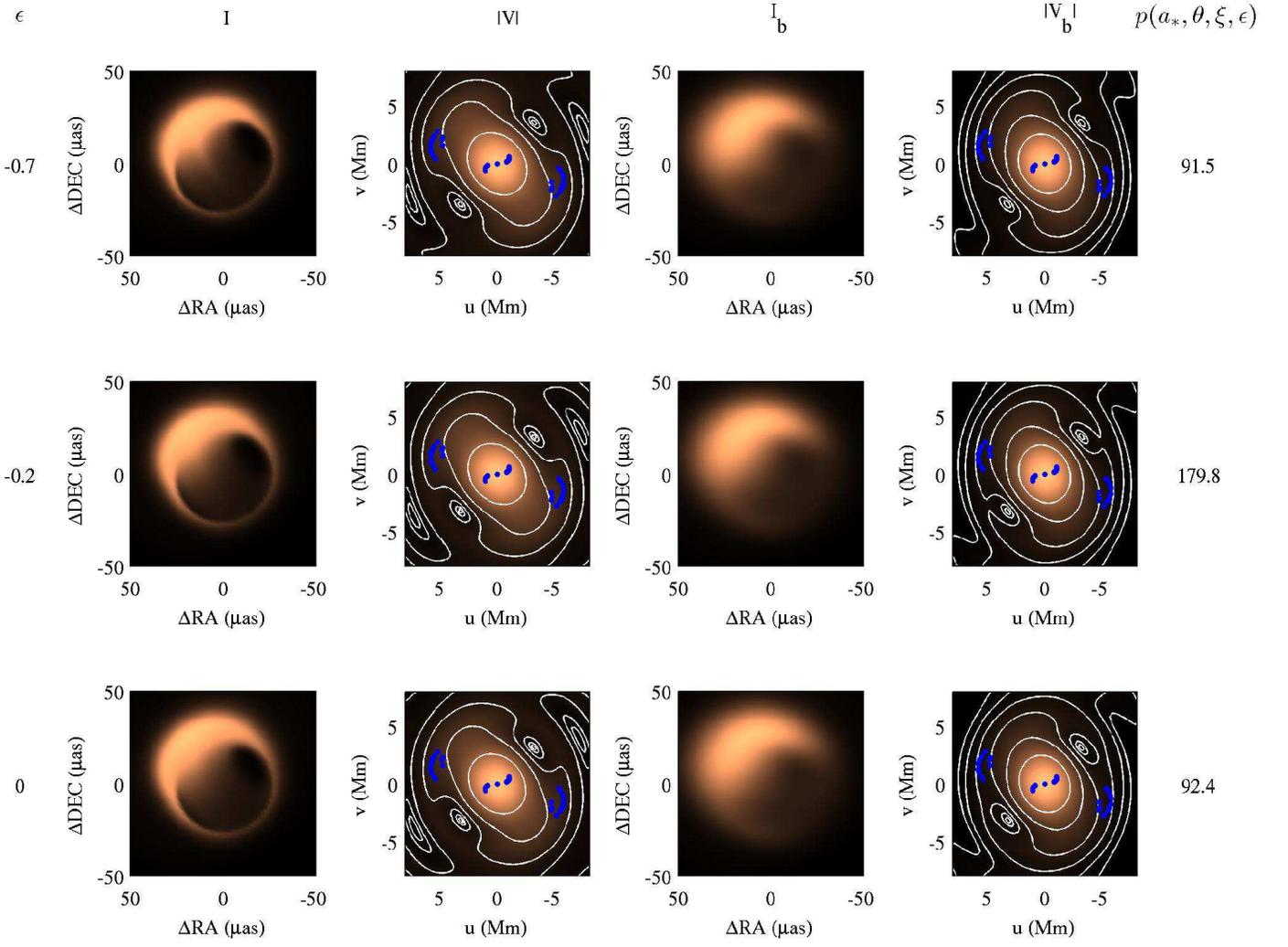}
\end{center}
\caption{Images and visibility magnitudes for the most probable model
  (middle row) and half as probable models obtained by varying the
  quadrupolar perturbation, $\epsilon$,  (top and bottom rows).
  Panels in columns 1-2 and 3-4, respectively, correspond to the intrinsic and
  scatter-broadened images.  The contours in the visibility magnitude
  plots are spaced logarithmically, in steps of factors of 2 and
  normalized to the maximum value for the best-fit model.  In all
  plots the color scale is linear, with the same normalization.
  Finally, values of $\epsilon$ 
  are shown on the left and the associated posterior probability on
  the right.  For reference the points in the $u$-$v$ plane at which
  EHT data exists are shown by the blue points in the visibility
  magnitude plots.
 }\label{fig:IVIV}
\end{figure*}

The fine degree of the distinctions between images being made by the
present EHT data are apparent in Figure \ref{fig:IVIV}.  There the
most probable image (middle row) is compared to images with identical
black hole spin but different quadrupolar perturbation parameter such
that the posterior probability density is reduced by roughly half.
As anticipated the variations in the black hole silhouette
induce corresponding variations in the visibilities that persist after
Galactic electron scattering broadens the image.  The electron scatter
broadening substantially reduces the signal associated with variations
on small scales, suggesting that the main constraining power arises
due to the change in the ISCO location and shadow shape on large
scales.  However, the over-all flux normalization, fixed
observationally by the spectral data, couples the dynamical impact of
the spacetime on small scales (long baseline length) to those on large
scales (small baseline length).  Hence it remains difficult to
unambiguously identify the primary source of the constraint.
Nevertheless, it is clear that structural differences do exist on
baselines already accessible by EHT observations.

\begin{figure*}
\begin{center}
\includegraphics[width=0.33\textwidth]{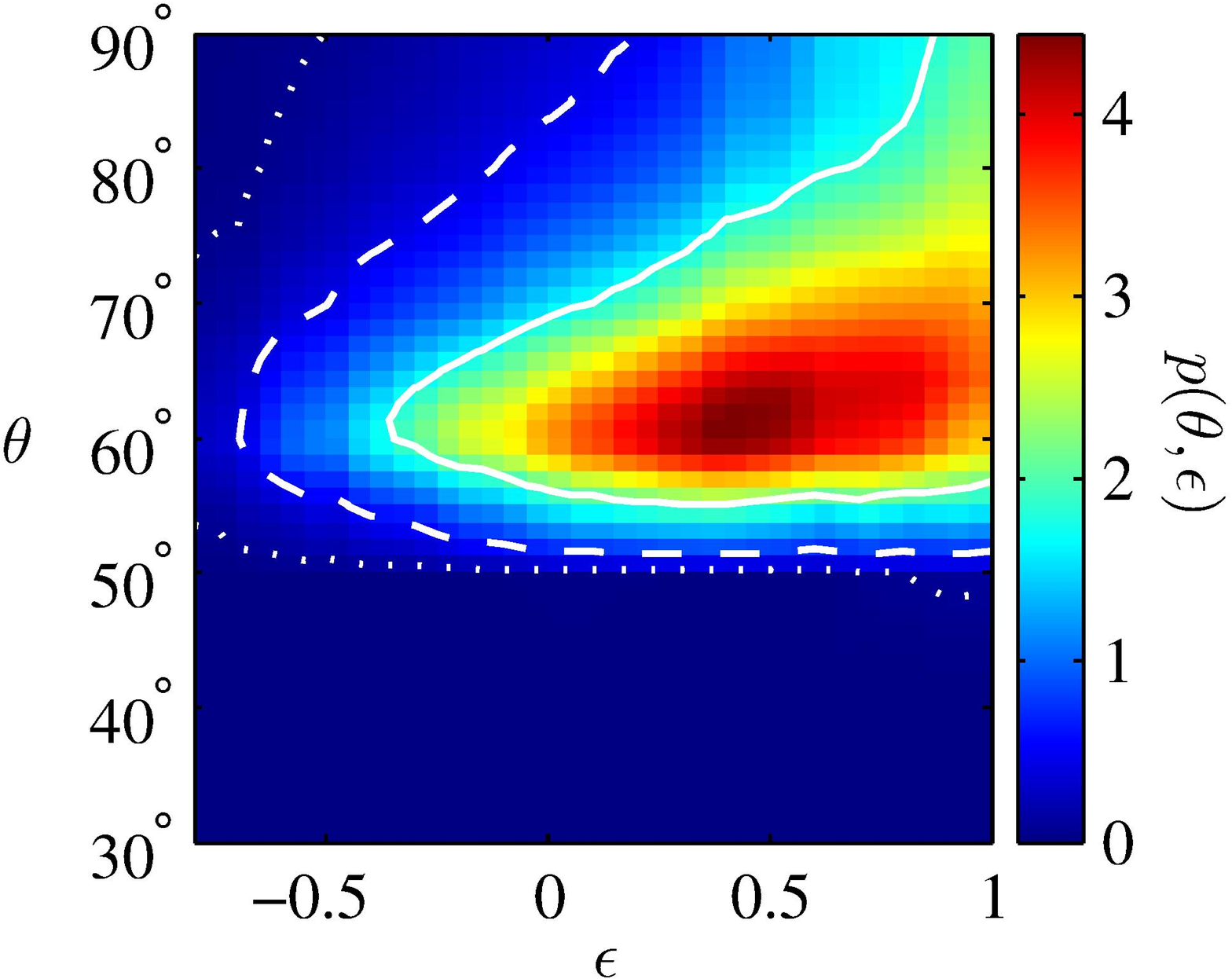}
\includegraphics[width=0.33\textwidth]{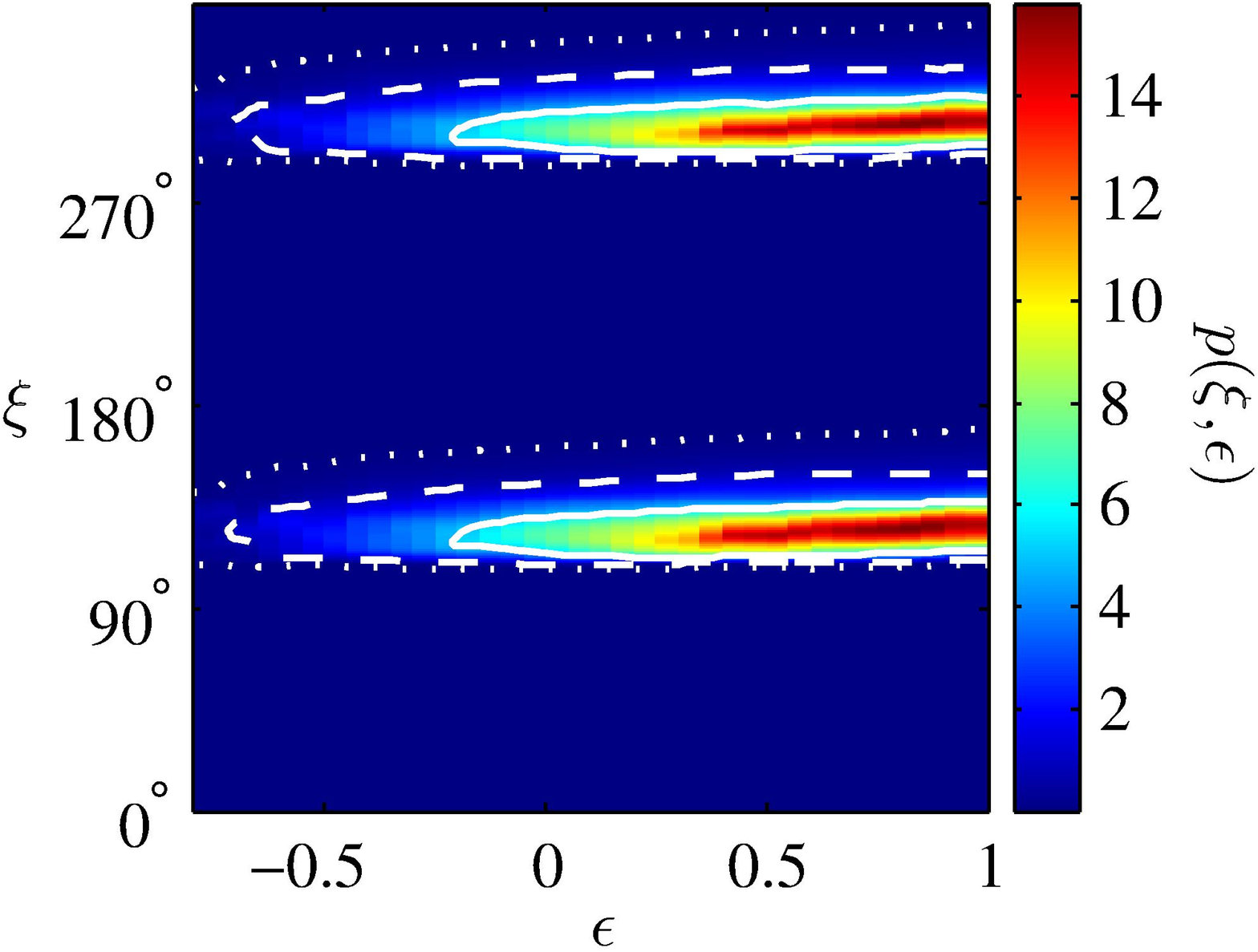}
\includegraphics[width=0.33\textwidth]{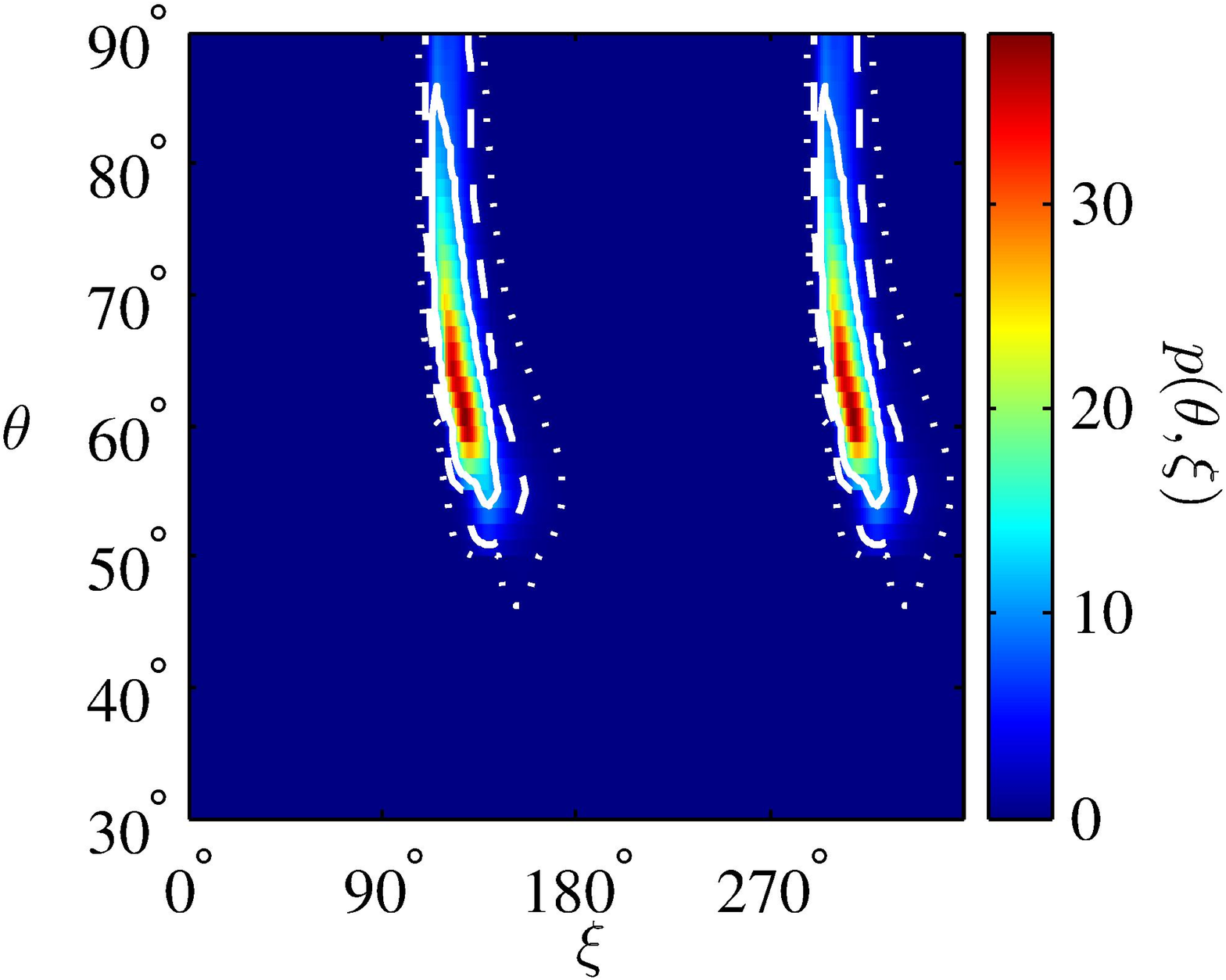}
\end{center}
\caption{Posterior probability densities as functions of 
  inclination and quadrupolar perturbation parameter (left), position
  angle and quadrupolar perturbation parameter (center), and
  inclination and position angle (right).  In all cases the posterior
  probabilities are marginalized over all other parameters.  For
  reference the 1$\sigma$, 2$\sigma$, and 3$\sigma$ contours are shown
  by the solid, dashed, and dotted white lines, respectively.
}\label{fig:dir}
\end{figure*}

\begin{figure}
\begin{center}
\includegraphics[width=0.8\columnwidth]{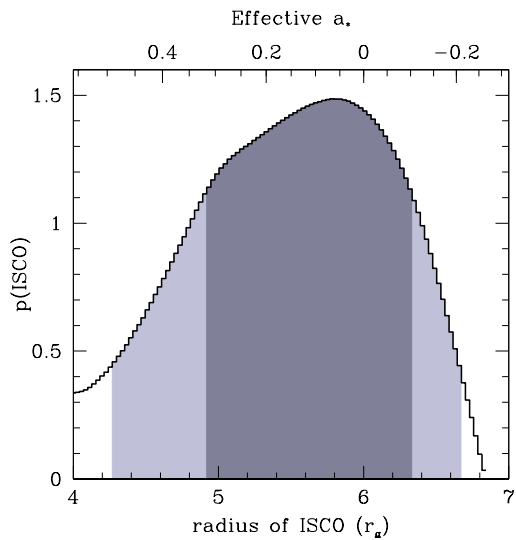}
\end{center}
\caption{Posterior probability density as a function of the ISCO
  radius (in $\Rg$), marginalized over all other quantities.  Dark and
  light gray shaded regions denote the 1$\sigma$ and 2$\sigma$
  intervals, respectively.  For reference, the inferred dimensionless
  spin magnitude for a Kerr black hole is shown on the top axis.  The
  inclination and spin orientation angles depend only very weakly on
  the deviation parameter.} \label{fig:pISCO} 
\end{figure}

\begin{figure}
\begin{center}
\includegraphics[width=\columnwidth]{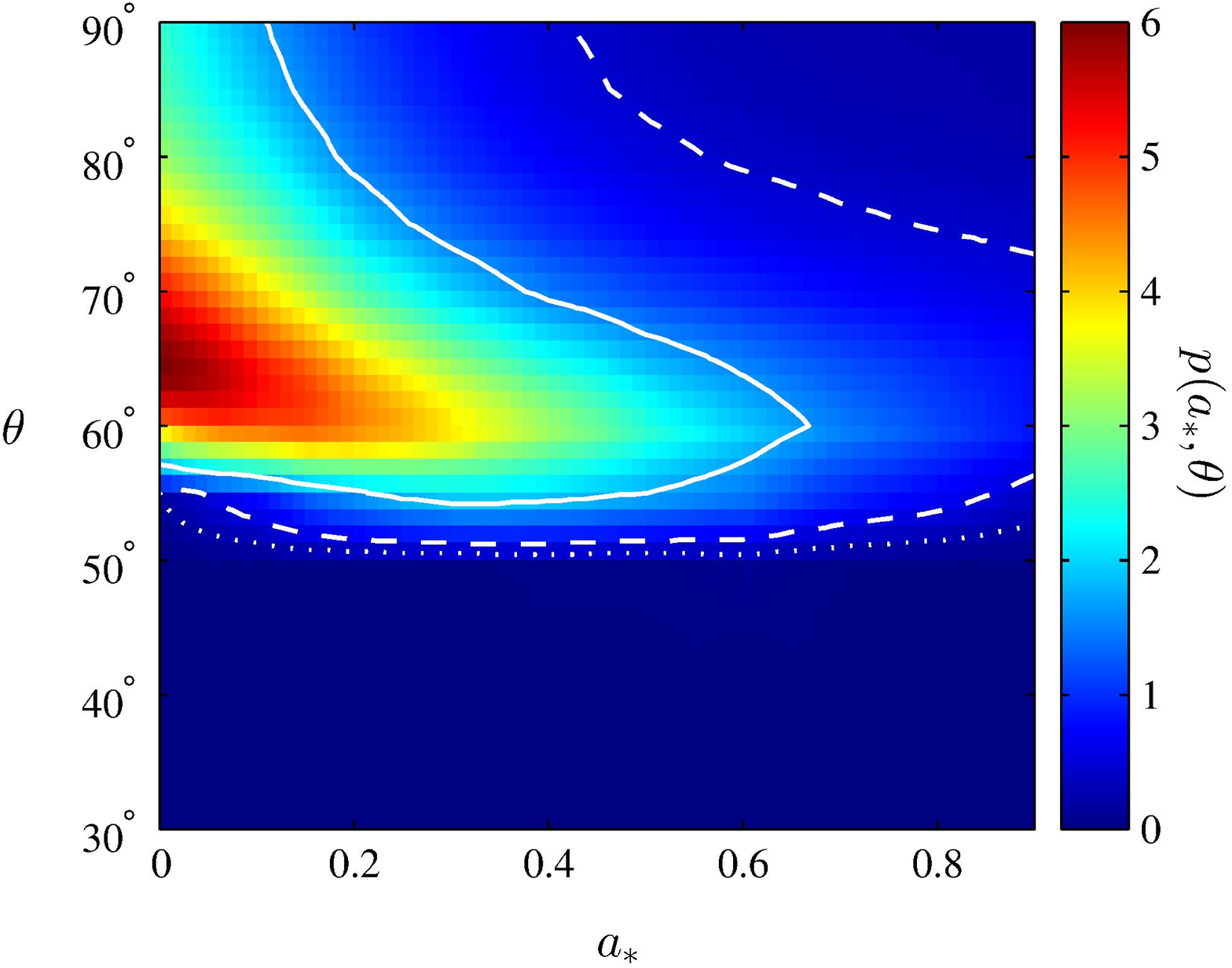}
\end{center}
\caption{Posterior probability density as a function of dimensionless
  spin magnitude and inclination, marginalized over all other
  quantities.  Solid, dashed, and dotted white lines show the
  1$\sigma$, 2$\sigma$, and 3$\sigma$ confidence
  contours.}\label{fig:path}
\end{figure}

\subsection{Implications for Previous Studies} \label{sec:Imps}

A key outstanding issue regarding the analysis in \citet{Brod_etal:11} is
robustness: it is unclear how robust the conclusions for spin
magnitude and direction are to variations in the disk model and/or
spacetime structure.  Here we attempt to address this issue, using the
quasi-Kerr metric as a proxy for just such a modification, and
assessing the associated sensitivity of the spin parameter estimates.
Of particular relevance is the fact that non-zero values of the
parameter $\epsilon$ probe larger ISCO radii, approximating the
retrograde disk scenario, which heretofore has been neglected on
dynamical grounds.

Despite the above concerns, it appears that the estimates of the spin
orientation are reasonably robust.  As seen in the left panel of
Figure \ref{fig:dir}, the inclination is only weakly correlated with
$\epsilon$.  Most importantly, the lower limit upon the
inclination, immediately relevant for the ability to image a
silhouette and the potential for dynamical lensing signals, is
independent of the quadrupolar perturbation.  This is independent of
the prior placed upon $\theta$, i.e., choosing a flat prior on
$\theta$ instead of the isotropic prior employed does not make a
significant difference in the lower-limit upon the inclination.
The position angle, center panel of Figure \ref{fig:dir}, is also
nearly independent of $\epsilon$, with similar conclusions
holding.  The result is that the posterior probability distribution of
the spin orientation (right panel of Figure \ref{fig:dir}) is tightly
constrained despite the additional freedom afforded by the spacetime
perturbation.  Hence, the inferred spin orientation (or in the case of
vanishing spin, disk orientation) is robust to the sorts of
perturbations considered here.

The implied impact upon the spin magnitude depends somewhat upon the
precise quantity being measured.  In Figure \ref{fig:pall}, the
high-probability density regions lie roughly along lines of fixed
ISCO, implying that the primary quantity being measured is not the
spin magnitude per se, but the dynamics  of spacetime orbits and their
subsequent effect upon the accretion flow.  As a result, $a$ and
$\epsilon$ appear substantially correlated, while the ISCO is
relatively robustly determined.  As seen in Figure \ref{fig:pISCO}, 
the radius of the ISCO is localized near $6 \Rg$, falling rapidly at
both substantially smaller and larger values.  For reference, the
associated spin parameter for a Kerr spacetime is shown on the top
axis.  Assuming that the variation of the ISCO is the dominant way in
which spin enters into the EHT constraints, it implies a spin
magnitude limit of $ a_*=0.05^{+0.27}_{-0.15}$, which is comparable to
the marginalized limits obtained in \citet{Brod_etal:11}.

The rapid decline in probability density at ISCO larger $6 \Rg$
provides an indirect justification for the neglect of retrograde
accretion flows.  These were previously ignored due to the dramatic
dynamical consequences associated with the interplay between the
Lense-Thirring torques and the orbital motion \citep{Frag_etal:07},
leading to concerns that stationary models of retrograde accretion
flows were intrinsically inconsistent.  However, insofar as the
primary impact of the retrograde orbital motion is to increase the
ISCO, Figure \ref{fig:pISCO} argues empirically against retrograde
accretion flow models, and thus for low and positive spin values.

Despite the strong correlation between $a$ and $\epsilon$, upon
marginalization over the quadrupolar perturbation, the inferred spin
magnitude is also reasonably robust.  This may be seen explicitly in
Figure \ref{fig:path}, which shows the posterior probability
distribution of spin magnitude and inclination, from which we find
$ a_*=0^{+0.7}$.  This is directly comparable to Figure 4b in
\citet{Brod_etal:11b}, to which it is similar.   While it does exhibit
a modest reduction in the small $ a_*$-$\theta$ correlation, likely
due to the limiting of our present analysis to $ a_*<0.9$ and
artifacts associated with the adaptive refinement scheme, the
high-probability structures are completely consistent.  

Quantitatively, this depends on the prior upon $\epsilon$
that is assumed.  The flat prior on $\epsilon$ employed here
combined with the restriction upon the ISCO we have adopted
effectively results in a down-weighting of $\epsilon\lesssim0.3$, 
for which portions of the $a$-$\epsilon$ parameter space are
excluded.  Adopting a uniform prior upon $\epsilon$, i.e.,
choosing the prior such that after marginalizing the prior over $a$
the resulting prior probability density in $\epsilon$ is flat,
up-weights small $\epsilon$, and thus small spins.  Thus, the
posterior probability distribution shown in Figure \ref{fig:path}
represents a rather pessimistic estimate of the limits upon the spin
magnitude.  Its similarity with previous analyses suggests that
efforts to constrain the spin magnitude are insensitive to even large
spacetime perturbations.

\subsection{Limits upon Deviations from the Kerr Metric} \label{sec:Lims}

\begin{figure}
\begin{center}
\includegraphics[width=\columnwidth]{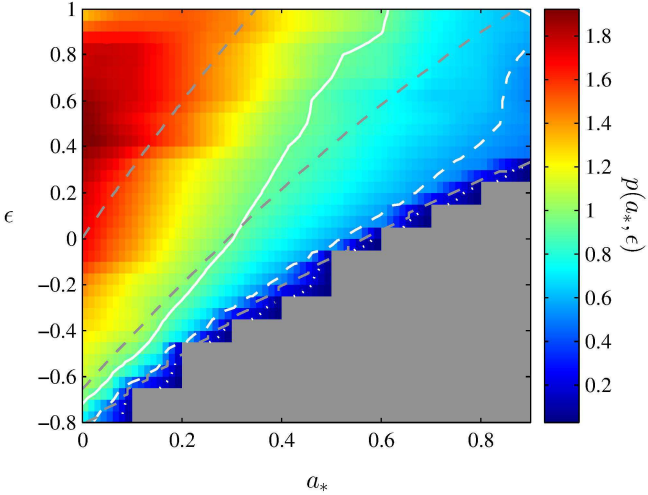}
\end{center}
\caption{Posterior probability density as a function of dimensionless
  spin magnitude and quadrupolar perturbation, marginalized
  over all other quantities.  The solid, dashed, and dotted lines shown the
  1$\sigma$, 2$\sigma$, and 3$\sigma$ confidence regions.  Lines of
  constant ISCO are shown by the dashed gray lines, corresponding to
  $6\Rg$, $5\Rg$, and $4\Rg$ from top to bottom.  The grayed
  region in the lower right are parameter values excluded by the
  condition that the ISCO lie outside of $4\Rg$.}\label{fig:pae}
\end{figure}

While the high-probability density regions lie roughly along contours
of constant ISCO radius, the probability density is not uniformly
distributed in $\epsilon$.  Indeed, this is partly responsible for the
robustness of the spin magnitude limits found described in the
preceding section.  It also suggests that it may be possible to
constrain $\epsilon$ from the current data.

The posterior probability distribution in spin magnitude and
quadrupolar perturbation is shown in Figure \ref{fig:pae}.  While a
modest concentration does exist at positive $\epsilon$ and
small $a$, the 2$\sigma$ confidence region extends essentially over
the entire $a$-$\epsilon$ parameter space considered.  Thus, as
yet, there is little that can be said regarding the value of
$\epsilon$ in a prior-independent manner.  Nevertheless, the strong
constraints apparent in the full parameter space (Figure \ref{fig:pall})
suggest that such limits will be forthcoming.  We will
address the strength of future limits from the EHT elsewhere.

\section{Conclusions} \label{sec:C}

We have constructed the first simulated images of a RIAF around a
supermassive black hole described by a quasi-Kerr metric, which
deviates from the Kerr metric in that it contains an independent
quadrupole moment.  Millimeter-wavelength VLBI observations are
capable of constraining the magnitude of the quadrupolar deviations
from the Kerr metric, and therefore violations of the no-hair theorem.
This remains true after folding in the impact of realistic accretion
flow and radiative transfer models.  The constraining capability
arises due to both the modification of the null geodesic structure of
the spacetime, and therefore the shape of the black hole shadow, as
well as the impact upon the dynamics of the accretion flow itself.  It
remains unclear to what degree these may be explicitly separated,
though current EHT observations are already capable of distinguishing
surprisingly similar images.  Unfortunately, the current published EHT
data does not place a strong limit upon the quadrupole deviation,
$\epsilon$, yet. We will address the strength of up-coming
observations elsewhere.

Despite the introduction of an additional model parameter, the result
for the black hole spin as inferred from previous studies is robust.  This
is especially true for the orientation, which is largely insensitive
to the quadrupolar distortion.  Even with the additional freedom, the
best fit inclination and position angles are
$\theta={65^\circ}^{+21^\circ}_{-11^\circ}$ and 
$\xi={127^\circ}^{+17^\circ}_{-14^\circ}$ (up to the $180^\circ$
degeneracy), consistent with previously obtained values.  This
suggests that the orientation of the spin (or accretion flow angular
momentum) is well constrained despite the substantial astrophysical
uncertainties that remain \citep[see][for a discussion]{Psal_etal:2013}.

The spin magnitude and quadrupolar perturbation are correlated such
that the ISCO is limited to $(5.8^{+0.7}_{-1.1})\Rg$.  Within the
context of the Kerr metric, this argues against the need to consider
retrograde accretion flows, previously disregarded on dynamical
grounds.  If translated into a limit upon the spin magnitude
using the ISCO-spin relationship of the Kerr spacetime, it implies
$ a_*=0.05^{+0.27}_{-0.15}$.

Despite this correlation, the posterior probability
distribution of the spin magnitude is quite similar to that found in
previous analyses of the EHT data, implying $ a_*=0^{+0.7}$.  While
this conclusion depends quantitatively upon the priors assumed for the
quadrupolar perturbation, it is qualitatively true for all reasonable
choices.

Here we restricted our analysis to spacetimes for which the ISCO lies
at a radius greater or equal to $4\Rg$ in order to avoid potential
pathologies of the quasi-Kerr arising at smaller radii. Since Sgr~A*
appears to be only slowly-spinning, this may be sufficient for this
source.  Typically, however, the most dramatic effects due to
deviations from the Kerr metric manifest at small radii, accessible to
accretion flows around rapidly-spinning black holes.  Generating
images of RIAFs about these black holes requires the use of Kerr-like
metrics that are free of pathological regions
\citep[e.g.,][]{JP11_PRD}.  We will present such images in
future work.

\begin{appendix}

\section{Analytical Flux Normalization Likelihood Marginalization/Maximization} \label{app:Fnorm}
Each observational epoch has an unknown flux normalization, assumed to
arise from moderate fluctuations in the accretion rate which to lowest
order just rescale the intensities across an image.  Here we obtain
analytical expressions in the presence of Gaussian errors and a flat
prior upon this normalization, $V_{00}$, for the marginalized and
maximized likelihoods.

In practice, for each observational epoch $E$, we have a set of
observed visibilities, $V_{j,E}$, at the points $(u_j,v_j)$, with some
error estimates $\sigma_{j,E}$.  Similarly, we have a set of model
visibilities $V_{00,E}\hat{V}_j(\bmath{p})$ where $\bmath{p}$ are the
various parameters that define the image model beyond the
observational epoch-dependent flux normalizations $V_{00,E}$.  Then
the associated likelihood is
\begin{equation}
L(V_{00,1},V_{00,2},\dots;\bmath{p})
=
N \exp\left[ 
  \sum_{E,j}
  \frac{\left(V_{j,E}-V_{00,E}\hat{V}_{j}\right)^2}{2\sigma_{j,E}^2} \right]\,,
\end{equation}
where $N$ is a fixed normalization.  Without loss of generality, we
can define $y_{j,E}\equiv V_{j,E}/\sigma_{j,E}$ and 
$\hat{y}_{j}\equiv \hat{V}_{j}/\sigma_{j,E}$, and thus treat only
unit-normalized distributions, i.e., 
\begin{equation}
L = N \exp\left[ \frac{1}{2}\sum_{E,j}
  \left(y_{j,E}-V_{00,E}\hat{y}_j\right)^2 \right]\,.
\end{equation}
In the above and henceforth, for simplicity we will suppress all
functional dependencies.

This is maximized relative to the flux normalization when 
\begin{equation}
\frac{\partial L}{\partial V_{00,E}} 
= 
\left[\sum_j \left(y_{j,E}-V_{00,E} \hat{y}_j\right) \hat{y}_j\right] L
=
0\,,
\end{equation}
and thus, we obtain the standard result
\begin{equation}
V_{00,E}^M = \frac{\sum_J y_{j,E} \hat{y}_j}{\sum_J \hat{y}_j^2}\,.
\end{equation}
Inserting this back into the likelihood and simplifying gives the
desired analytical form for the maximum likelihood:
\begin{equation}
L^M = N \prod_E \exp \left[
\frac{
  \left(\sum_j y_{j,E}^2\right)
  \left(\sum_j \hat{y}_j^2\right)
  -
  \left(\sum_j y_{j,E} \hat{y}_j\right)^2
}{
  2 \sum_j \hat{y}_j^2
}
\right]\,.
\end{equation}
\mbox{}

On the other hand, the marginalized likelihood is defined by
\begin{equation}
\begin{aligned}
\bar{L} 
&= 
\int dV_{00,1} dV_{00,2} \dots \pi(V_{00,1})\pi(V_{00,2})\dots L\\
&=
L^M
\prod_E \int dV_{00,E}\\
&\qquad\times
\Pi(V_{00,E})
\exp
\left[
-
\left(\sum_j \frac{\hat{y}_j^2}{2}\right)
\left(V_{00,E} - V^M_{00,E}\right)^2
\right]\,,
\end{aligned}
\end{equation}
where $\Pi(V_{00,E})$ is the prior on $V_{00,E}$.  If $\Pi(V_{00,E})$
varies slowly over variations in $V_{00,E}$ that are much wider than
the peak of the exponential in the final integral, we may approximate
it as flat and extending from $V_{00,E}=-\infty$ to $\infty$.  Thus,
the average value of $V_{00,E}$ is simply $V^M_{00,E}$, and the the
resulting marginalized likelihood is
\begin{equation}
\bar{L}
=
L^M \prod_E \left(\frac{2\pi}{\sum_j \hat{y}_j^2}\right)^{1/2} \Pi(V_{00,E}^M)\,,
\end{equation}
where $\Pi(V_{00,E}^M)$ is simply a constant, and may be neglected in
further parameter estimation.

\section{Marginalized Likelihoods and Posterior Probability Construction} \label{app:MLP}
The desired analysis product is essentially a posterior probability
distribution, marginalized over various nuisance parameters (in the
presence context, the epoch-dependent flux normalizations).  Here we
show that these are trivially related to the marginalized likelihoods.

Consider a model with a set of $n$ interesting parameters, $\p$, and $m$
nuisance parameters, $\q$, for which we have likelihoods $L(\p,\q)$ with
priors $\Pi(\p,\q)$.  Then the posterior probability density is given
by Bayes' theorem by,
\begin{equation}
p(\p,\q)
=
\frac{L(\p,\q) \Pi(\p,\q)}{\int d^n\!p\,d^m\!q\, L(\p,\q) \Pi(\p,\q)}\,.
\end{equation}
After marginalizing over the nuisance parameters, we obtain the
desired probability density,
\begin{equation}
p(\p)
=
\int d^m\!q\,p(\p,\q)
=
\frac{\int d^m\!q\, L(\p,\q) \Pi(\p,\q)}{\int d^n\!p\,d^m\!q\, L(\p,\q) \Pi(\p,\q)}\,.
\end{equation}

If the priors upon the physical and nuisance parameters are separable,
i.e., $\Pi(\p,\q)=\Pi_p(\p)\Pi_q(\q)$, then this may be written as
\begin{equation}
p(\p)
=
\frac{\bar{L}(\p) \Pi_p(\p)}{\int d^n\!p\, \bar{L}(\p) \Pi_p(\p)}
\,.
\end{equation}
where
\begin{equation}
\bar{L}(\p)\equiv\int d^m\!q\, L(\p,\q) \Pi_q(\q)
\end{equation}
are the marginalized likelihoods.  Thus the marginalization over the
nuisance parameters may be performed during the construction of the
likelihoods themselves, as in Appendix \ref{app:Fnorm}.

\end{appendix}

\acknowledgments
A.E.B.~receives financial support from Perimeter Institute for
Theoretical Physics and the Natural Sciences and Engineering Research
Council of Canada through a Discovery Grant. T.J.~was supported by a
CITA National Fellowship at the University of Waterloo and in part by
Perimeter Institute for Theoretical Physics. Research at Perimeter
Institute is supported by the Government of Canada through Industry
Canada and by the Province of Ontario through the Ministry of Research
and Innovation. A.L.~was supported in part by NSF grants AST-0907890,
AST-0807843, and AST-0905844, and NASA grants NNX08AL43G and
NNA09DB30A. D.P.~was supported by an NSF CAREER award NSF 0746549. 

\bibliography{gcpe2.bib}
\bibliographystyle{apj}

\end{document}